\documentclass[modern,tighten]{aastex62}
\usepackage[utf8]{inputenc}
\usepackage{graphicx}

\def\be{\begin{equation}}
\def\ee{\end{equation}}
\def\ba{\begin{eqnarray}}
\def\ea{\end{eqnarray}}

\newcommand{\msun}{\ifmmode\mbox{M}_{\odot}\else$\mbox{M}_{\odot}$\fi}
\newcommand{\rsun}{\ifmmode\mbox{R}_{\odot}\else$\mbox{R}_{\odot}$\fi}
\newcommand{\degrees}{\ifmmode^{\circ}\else$^{\circ}$\fi}
\newcommand{\degree}{\ifmmode^{\circ}\else$^{\circ}$\fi}
\newcommand{\amin}{\ifmmode^{\prime}\else$^{\prime}$\fi}
\newcommand{\asec}

\begin{document}

\title{\Large High-Precision X-ray Timing of Three Millisecond Pulsars with \textit{NICER}: Stability Estimates and Comparison with Radio}

\author[0000-0003-1226-0793]{J.~S.~Deneva}
\affiliation{George Mason University, resident at the Naval Research Laboratory, Washington, DC 20375, USA}
\author[0000-0002-5297-5278]{P.~S.~Ray}
\affiliation{U.S. Naval Research Laboratory, Washington, DC 20375, USA}
\author{A.~Lommen}
\affiliation{Haverford College, Haverford, PA 19041, USA}
\author[0000-0001-5799-9714]{S.~M.~Ransom}
\affiliation{National Radio Astronomy Observatory, Charlottesville, VA, USA}
\author[0000-0002-9870-2742]{S.~Bogdanov}
\affil{Columbia Astrophysics Laboratory, Columbia University, New York, NY 10027, USA}
\author{M.~Kerr}
\affiliation{U.S. Naval Research Laboratory, Washington, DC 20375, USA}
\author{K.~S.~Wood}
\affiliation{Praxis, resident at the Naval Research Laboratory, Washington, DC 20375, USA}
\author{Z.~Arzoumanian}
\affiliation{X-Ray Astrophysics Laboratory, NASA Goddard Space Flight Center, Greenbelt, MD 20771, USA}
\author{K.~Black}
\affiliation{Goddard Space Flight Center, Greenbelt, MD 20771, USA}
\author{J.~Doty}
\affiliation{Noqsi Aerospace Ltd, Billerica, MA 01821, USA}
\author{K.~C.~Gendreau}
\affiliation{X-Ray Astrophysics Laboratory, NASA Goddard Space Flight Center, Greenbelt, MD 20771, USA}
\author[0000-0002-6449-106X]{S. Guillot}
\affiliation{IRAP, CNRS, 9 avenue du Colonel Roche, BP 44346, F-31028 Toulouse Cedex 4, France}
\affiliation{Universit\'{e} de Toulouse, CNES, UPS-OMP, F-31028 Toulouse, France.}
\author{A.~Harding}
\affiliation{Goddard Space Flight Center, Greenbelt, MD 20771, USA}
\author{N.~Lewandowska}
\affiliation{Department of Physics and Astronomy, West Virginia University, P.O. Box 6315, Morgantown, WV 26506, USA}
\affiliation{Center for Gravitational Waves and Cosmology, West Virginia University, Chestnut Ridge Research Building, Morgantown, WV 26505, USA}
\author{C.~Malacaria}
\affiliation{ST12 Astrophysics Branch, NASA Marshall Space Flight Center, Huntsville, AL 35812, USA}
\affiliation{Universities Space Research Association, NSSTC, Huntsville, AL 35805, USA}
\author{C.~B.~Markwardt}
\affiliation{Goddard Space Flight Center, Greenbelt, MD 20771, USA}
\author{S.~Price}
\affiliation{Goddard Space Flight Center, Greenbelt, MD 20771, USA}
\author{L.~Winternitz}
\affiliation{Goddard Space Flight Center, Greenbelt, MD 20771, USA}
\author[0000-0002-4013-5650]{M.~T.~Wolff}
\affiliation{U.S. Naval Research Laboratory, Washington, DC 20375, USA}
\author{L.~Guillemot}
\affiliation{Laboratoire de Physique et Chimie de l'Environnement et de l'Espace, LPC2E, CNRS-Universit\'{e} d'Orl\'{e}ans, F-45071 Orl\'{e}ans, France}
\affiliation{Station de Radioastronomie de Nan\c{c}ay, Observatoire de Paris, CNRS/INSU, F-18330 Nan\c{c}ay, France}
\author{I.~Cognard}
\affiliation{Laboratoire de Physique et Chimie de l'Environnement et de l'Espace, LPC2E, CNRS-Universit\'{e} d'Orl\'{e}ans, F-45071 Orl\'{e}ans, France}
\affiliation{Station de Radioastronomie de Nan\c{c}ay, Observatoire de Paris, CNRS/INSU, F-18330 Nan\c{c}ay, France}
\author[0000-0003-2745-753X]{P.\,T.\,Baker}
\affiliation{Department of Physics and Astronomy, West Virginia University, P.O. Box 6315, Morgantown, WV 26506, USA}
\affiliation{Center for Gravitational Waves and Cosmology, West Virginia University, Chestnut Ridge Research Building, Morgantown, WV 26505, USA}
\author{H.\,Blumer}
\affiliation{Department of Physics and Astronomy, West Virginia University, P.O. Box 6315, Morgantown, WV 26506, USA}
\affiliation{Center for Gravitational Waves and Cosmology, West Virginia University, Chestnut Ridge Research Building, Morgantown, WV 26505, USA}
\author{P.\,R.\,Brook}
\affiliation{Department of Physics and Astronomy, West Virginia University, P.O. Box 6315, Morgantown, WV 26506, USA}
\affiliation{Center for Gravitational Waves and Cosmology, West Virginia University, Chestnut Ridge Research Building, Morgantown, WV 26505, USA}
\author{H.\,T.\,Cromartie}
\affiliation{University of Virginia, Department of Astronomy, P.O. Box 400325, Charlottesville, VA 22904, USA}
\author{P.\,B.\,Demorest}
\affiliation{National Radio Astronomy Observatory, 1003 Lopezville Rd., Socorro, NM 87801, USA}
\author{M.\,E.\,DeCesar}
\affiliation{Department of Physics, Lafayette College, Easton, PA 18042, USA}
\author[0000-0001-8885-6388]{T.\,Dolch}
\affiliation{Department of Physics, Hillsdale College, 33 E. College Street, Hillsdale, Michigan 49242, USA}
\author{J.\,A.\,Ellis}
\affiliation{Infinia ML, 202 Rigsbee Avenue, Durham NC, 27701}
\author{R.\,D.\,Ferdman}
\affiliation{School of Chemistry, University of East Anglia, Norwich, NR4 7TJ, United Kingdom}
\author{E.\,C.\,Ferrara}
\affiliation{NASA Goddard Space Flight Center, Greenbelt, MD 20771, USA}
\author[0000-0001-8384-5049]{E.\,Fonseca}
\affiliation{Department of Physics, McGill University, 3600  University St., Montreal, QC H3A 2T8, Canada}
\author{N.\,Garver-Daniels}
\affiliation{Department of Physics and Astronomy, West Virginia University, P.O. Box 6315, Morgantown, WV 26506, USA}
\affiliation{Center for Gravitational Waves and Cosmology, West Virginia University, Chestnut Ridge Research Building, Morgantown, WV 26505, USA}
\author{P.\,A.\,Gentile}
\affiliation{Department of Physics and Astronomy, West Virginia University, P.O. Box 6315, Morgantown, WV 26506, USA}
\affiliation{Center for Gravitational Waves and Cosmology, West Virginia University, Chestnut Ridge Research Building, Morgantown, WV 26505, USA}
\author{M.\,L.\,Jones}
\affiliation{Department of Physics and Astronomy, West Virginia University, P.O. Box 6315, Morgantown, WV 26506, USA}
\affiliation{Center for Gravitational Waves and Cosmology, West Virginia University, Chestnut Ridge Research Building, Morgantown, WV 26505, USA}
\author[0000-0003-0721-651X]{M.\,T.\,Lam}
\affiliation{Department of Physics and Astronomy, West Virginia University, P.O. Box 6315, Morgantown, WV 26506, USA}
\affiliation{Center for Gravitational Waves and Cosmology, West Virginia University, Chestnut Ridge Research Building, Morgantown, WV 26505, USA}
\author{D.\,R.\,Lorimer}
\affiliation{Department of Physics and Astronomy, West Virginia University, P.O. Box 6315, Morgantown, WV 26506, USA}
\affiliation{Center for Gravitational Waves and Cosmology, West Virginia University, Chestnut Ridge Research Building, Morgantown, WV 26505, USA}
\author{R.\,S.\,Lynch}
\affiliation{Green Bank Observatory, P.O. Box 2, Green Bank, WV 24944, USA}
\author[0000-0001-7697-7422]{M.\,A.\,McLaughlin}
\affiliation{Department of Physics and Astronomy, West Virginia University, P.O. Box 6315, Morgantown, WV 26506, USA}
\affiliation{Center for Gravitational Waves and Cosmology, West Virginia University, Chestnut Ridge Research Building, Morgantown, WV 26505, USA}
\author[0000-0002-3616-5160]{C.\,Ng}
\affiliation{Department of Physics and Astronomy, University of British Columbia, 6224 Agricultural Road, Vancouver, BC V6T 1Z1, Canada}
\affiliation{Dunlap Institute, University of Toronto, 50 St. George St., Toronto, ON M5S 3H4, Canada}
\author[0000-0002-6709-2566]{D.\,J.\,Nice}
\affiliation{Department of Physics, Lafayette College, Easton, PA 18042, USA}
\author[0000-0001-5465-2889]{T.\,T.\,Pennucci}
\affiliation{Hungarian Academy of Sciences MTA-ELTE Extragalactic Astrophysics Research Group, Institute of Physics, E\"{o}tv\"{o}s Lor\'{a}nd University, P\'{a}zm\'{a}ny P. s. 1/A, Budapest 1117, Hungary}
\author[0000-0002-6730-3298]{R.\,Spiewak}
\affiliation{Centre for Astrophysics and Supercomputing, Swinburne University of Technology, P.O. Box 218, Hawthorn, Victoria 3122, Australia}
\author[0000-0001-9784-8670]{I.\,H.\,Stairs}
\affiliation{Department of Physics and Astronomy, University of British Columbia, 6224 Agricultural Road, Vancouver, BC V6T 1Z1, Canada}
\author{K.\,Stovall}
\affiliation{National Radio Astronomy Observatory, 1003 Lopezville Rd., Socorro, NM 87801, USA}
\author{J.\,K.\,Swiggum}
\affiliation{Center for Gravitation, Cosmology and Astrophysics, Department of Physics, University of Wisconsin-Milwaukee, P.O. Box 413, Milwaukee, WI 53201, USA}
\author{S.\,J.\,Vigeland}
\affiliation{Center for Gravitation, Cosmology and Astrophysics, Department of Physics, University of Wisconsin-Milwaukee, P.O. Box 413, Milwaukee, WI 53201, USA}
\author{W.\,W.\,Zhu}
\affiliation{National Astronomical Observatories, Chinese Academy of Science, 20A Datun Road, Chaoyang District, Beijing 100012, China}

\begin{abstract}
The \textit{Neutron Star Interior Composition Explorer} (\textit{NICER}) is an 
X-ray astrophysics payload on
the International Space Station. It enables unprecedented high-precision timing of millisecond pulsars without the pulse broadening and delays due to dispersion and scattering within the interstellar medium that plague radio timing. We present initial timing results from a year of data on the millisecond pulsars PSR~B1937$+$21 and PSR~J0218$+$4232, and nine months of data on PSR~B1821$-$24. 
\textit{NICER} time-of-arrival uncertainties for the three pulsars are consistent with theoretical lower bounds and simulations based on their pulse shape templates and average source and background photon count rates. To estimate timing stability, we use the $\sigma_z$ measure, which is based on the average of the cubic coefficients of polynomial fits to subsets of timing residuals. So far we are achieving timing stabilities $\sigma_z \approx 3 \times 10^{-14}$ for PSR B1937+21 and on the order of $10^{-12}$ for PSRs B1821$-$24 and J0218+4232. Within the span of our \textit{NICER} data we do not yet see the characteristic break point in the slope of $\sigma_z$; detection of such a break would indicate that further improvement in the cumulative root-mean-square (RMS) timing residual is limited by timing noise. We see this break point in our comparison radio data sets for PSR~B1821$-$24 and PSR~B1937+21 on time scales of $> 2$~years. 
\end{abstract}


\section{Introduction}

The \emph{Neutron Star Interior Composition Explorer} (\textit{NICER}) is an X-ray instrument mounted on a movable arm on the outside of the International Space Station (ISS) and has been in operation since June 2017 \citep{Gendreau17}. \textit{NICER} was specifically designed to study the X-ray emissions of neutron stars (NSs). The main motivation is high-precision timing of X-ray emitting pulsars, constraining the mass-radius relation of NSs, and studying their high-energy emission mechanisms. High-precision X-ray timing with \textit{NICER} has already demonstrated how pulsars can be used for autonomous space navigation (\citealt{Mitchell18}, \citealt{Winternitz18}).

The detection of gravitational waves by LIGO \citep{Abbott16,Abbott17} has opened a new window for exploring astrophysical phenomena such as black hole and neutron star mergers. While LIGO is sensitive to the 10--100~Hz emission from the final inspiral of stellar-mass
compact objects, pulsar timing arrays (PTAs) use long-term millisecond pulsar (MSP) timing to attempt to detect \emph{nanohertz} gravitational waves from supermassive black hole binaries starting long before the system mergers. PTAs are omnidirectional, and their sensitivity improves as more observations are accumulated. The pulsars are always `on' so the experiment runs continuously, limited only by our ability to observe the pulsars. However, because PTAs have, to date, relied exclusively on radio timing, they must contend with the effects of the interstellar medium (ISM) on radio timing precision. The main ISM contributions to timing perturbations observed in PTA pulsars are from variable dispersion delays ($\propto \nu^{-2}$) and scattering delays ($\propto \nu^{-4}$) where $\nu$ is the frequency of radio emission. Most of the effect of dispersion can be corrected for during data processing. 
Scattering as well as variations in dispersion are stochastic processes that are difficult to model over long time scales \citep{Shannon17}. They are also difficult to completely disentangle from intrinsic timing noise or from the timing perturbations caused by nHz gravitational waves. 

X-ray timing observations of MSPs with \textit{NICER} can help separate propagation effects from intrinsic timing noise in PTA observations. X-rays are effectively at an infinite electromagnetic frequency compared to radio frequencies employed in PTA observations, and therefore \textit{NICER} observations of MSPs are immune to the timing effects of dispersion and scattering. 
\cite{Ray08} found that \textit{RXTE} TOAs of PSR~B1821$-$24 agree with a GBT radio timing solution to within the X-ray TOA error bars. However, the \textit{RXTE} observations produced only four TOAs within a year, which limits the time scale for estimating rotational instabilities. \textit{NICER} provides superior precision, weekly observations, and the opportunity to perform this analysis for more MSPs. 
  
Most MSPs are very faint in X-rays \citep[e.g.,][]{Webb04b} and many MSPs exhibit thermal X-ray pulsations that have substantially broader peaks compared to their radio counterparts \citep[e.g.,][]{Zavlin06,Bogdanov09}, so they are not well suited for precision timing. On the other hand, three MSPs are known to exhibit narrow non-thermal X-ray pulsations desirable for high-precision timing analyses: PSR~B1821$-$24, PSR~B1937+21, and PSR~J0218+4232. 

{\bf PSR B1821$-$24} (also known as PSR J1824$-$2452A) was the first radio MSP to be found in a globular cluster \citep{Lyne87}. 
Pulsed X-ray emission from this isolated $3.05$ ms rotator was first detected using \textit{ASCA} \citep{Saito97}; it features two remarkably narrow pulses per period with a high pulsed fraction ($\sim$80\%) and small duty cycle \citep{Rutledge04,Ray08}. These characteristics have made PSR B1821$-$24 the go-to pulsar for calibrating the absolute timing capabilities of X-ray observatories, including \textit{ASCA}\footnote{See \url{https://heasarc.gsfc.nasa.gov/docs/asca/newsletters/gis_time_assign5.html}}, \textit{Chandra}\footnote{See \url{http://cxc.harvard.edu/contrib/arots/time/CXOClock.pdf}}, and \textit{RXTE} \citep{Rots98}.

{\bf PSR B1937$+$21} (PSR J1939$+$2134), the first MSP to be discovered \citep{Backer82}, was found to be a pulsed X-ray source with \textit{ASCA} \citep{Takahashi01}. It was subsequently studied in more detail with \textit{BeppoSAX} \citep{Nicastro04}, \textit{RXTE} \citep{Guillemot12}, \textit{Chandra}, \textit{XMM-Newton} \citep{Ng14}, and \textit{NuSTAR} \citep{Gotthelf17}. The pulse profiles of PSR~B1821$-$24 and PSR~B1937$+$21 show a prominent and narrow main X-ray pulse and a much weaker interpulse.

{\bf PSR J0218$+$4232} is a 2.32 ms pulsar \citep{Navarro95} bound to a white dwarf companion in a two-day binary orbit. The pulsed X-ray emission from PSR J0218$+$4232 has previously been studied with \textit{BeppoSAX}   \citep{Mineo00},  \textit{Chandra} \citep{Kuiper02}, \textit{XMM-Newton} \citep{Webb04a}, and \textit{NuSTAR} \citep{Gotthelf17}, which revealed two moderately sharp pulses per period with a hard non-thermal spectrum.

In this paper, we use \textit{NICER} data to characterize the intrinsic rotational stabilities of PSR~B1821$-$24, PSR~B1937+21, and PSR~J0218+4232 and compare our findings with results from long-term radio observations from two PTA projects: the North American Nanohertz Observatory for Gravitational Waves
(NANOGrav\footnote{\url{http://nanograv.org}}) and the European Pulsar Timing Array (EPTA\footnote{\url{http://www.epta.eu.org}}). Section~\ref{sec-data} describes the selection criteria we apply to \textit{NICER} photons; Section~\ref{sec-timing} focuses on the pulsar timing procedure and differences between X-ray and radio timing; Section~\ref{sec-toa-errors} compares the uncertainties of \textit{NICER} pulse times-of-arrival (TOAs) with simulations and theoretical predictions; Section~\ref{sec-residuals} presents \textit{NICER} and radio timing residuals{\footnote{A {\em residual} in this case is the difference between the TOA and arrival time predicted by a model.}} for overlapping time spans; and Section~\ref{sec-sigmaz} discusses our findings about the rotational stability of each pulsar based on \textit{NICER} and radio data. 

\section{Data Selection}\label{sec-data}

\textit{NICER}'s X-ray Timing Instrument (XTI) comprises 56 paired X-ray optics and detectors (52 currently active) co-aligned to observe the same point on the sky. Each light path consists of a grazing-incidence X-ray ``concentrator'' and a silicon drift detector (for more detail, see \citealt{Gendreau16}, \citealt{Prigozhin16}). \textit{NICER} is sensitive to X-rays in the 0.2--12~keV range and has a peak collecting area of 1900~cm$^2$ at 1.5~keV. 

Energy deposition in the XTI's silicon detectors, from photons or charged particles, produces an amplified charge signal.  The charge signal is fed into two analog signal processing chains with different pulse shaping time constants: slow (465~ns peaking time) and fast (85~ns) \citep{Prigozhin16}. The slow chain is optimized for energy measurement, while the fast chain is optimized for time measurement. A preset threshold in each chain produces an electronic trigger that causes the event pulse height and timestamp to be sampled and digitized.  For events that trigger both chains, the fast-chain timestamp is reported; for events that trigger only the slow chain, the slow-chain timestamp is reported. In either case, the pulse height measured by the slow chain is reported.  In practice, X-rays below $\sim 1$~keV trigger only the slow chain; higher-energy X-rays trigger both chains. Energetic particles, as well as gamma rays produced as particles interact with the detector or surrounding shielding, can also cause both chains to trigger.

Event time stamps are referenced to \textit{NICER}'s on-board GPS receiver. They are affected by a bias (a fixed offset between a local clock and a reference clock) and an uncertainty; the latter includes contributions from time-stamping hardware accuracy, uncertainties in the lengths of cables within the instrument, and errors in the GPS receiver's realization of GPS time. Before launch, the time biases and uncertainties of the slow and fast chains were measured by laboratory equipment with calibrated biases. The fast chain timing uncertainty was determined to be 70~ns;  the difference between slow- and fast-chain timestamps was measured with 4~ns uncertainty.  Thus, for all practical purposes the two analog chains are identical in terms of timing uncertainty performance.  The slow- and fast-chain time biases were recorded separately for each \textit{NICER} detector (typically $\sim 250$~ns for the fast chain, $\sim 760$~ns for the slow chain, with $\sim 11$~ns variations between detectors).  These biases are corrected in standard processing of \textit{NICER} data (the {\tt nicertimecal} routine within the HEAsoft package; see below).  After correction, \textit{NICER} calibrated event timestamp values refer to the time that an X-ray or particle entered the detector aperture.

Particle-induced events typically have very high amplitudes and often occur far from the center of the detector (as the 25 mm$^2$ active area of the physical detector is larger than the 2 mm-diameter entrance aperture for X-rays, \citealt{Prigozhin16}). They are rejected using a combination of criteria based on amplitude and offset from the detector center \citep{Gendreau16}. The ratio ({\tt PI\_RATIO}) of amplitudes detected by the two signal chains for the same event is strongly dependent on the offset from the detector center and provides an effective way to filter out particle background. The recorded integer \texttt{PI} value is the gain-corrected energy of the photon in units of 10 eV. {\tt PI\_RATIO = PI/PI\_FAST}, where {\tt PI} refers to the slow chain, and we exclude events with ${\tt PI\_RATIO} > 1.1 + 120/{\tt PI}$. 


In this paper, we analyze data from 2017 July -- 2018 June using HEASoft 6.24\footnote{\url{https://heasarc.nasa.gov/lheasoft/}} and NICERDAS 2018-04-13\_V004. PSR~J0218+4232 was too close to the Sun to be observed after 30 March 2018, and therefore its effective date range is 2017 July -- 2018 March. PSR B1821$-$24 also goes behind the Sun for three months out of the year, roughly November -- February, so there is a gap in those data. 
We select Good Time Intervals (GTIs) using the following four criteria: the ISS is not within the South Atlantic Anomaly (SAA); \textit{NICER} is in tracking mode; \textit{NICER} is pointing within $0.015\degree$ of the source; and the source is at least 30\degree\ above the Earth's limb. We use only photons from within these GTIs.

\textit{NICER} observations are typically hundreds to thousands of seconds long, and there are often multiple observations of the same source with exposure times in this range per day. For data catalog purposes, observations from a given UTC day are grouped under an ``ObsID'' identifier and are downloaded, processed, and filtered together. Even after the filtering steps described above, there is still a considerable background left at low energies for some ObsIDs, often due to excessive counts in just a few detectors. This may be due to sunlight, either direct or reflected from the ISS structure, illuminating \textit{NICER} detectors unevenly, combined with differences in light sensitivity between detectors. In addition, enhanced background and flaring are often observed when the ISS is in regions of low geomagnetic cutoff rigidity (``polar horn'' regions), dependent on the current space weather conditions.

For each ObsID, we calculate the average count rate for each detector and exclude outlier detectors with count rates $> 3\sigma$ above the mean across detectors. We repeat this three times to obtain an average count rate per detector that is not contaminated by outliers. 
The excluded detectors as well as their total number for each ObsID differ between ObsIDs. For PSR~J0218+4232, 27 of 152 ObsIDs had 1--3 excluded detectors; for PSR~B1821$-$24, 15 of 146 ObsIDs had 1--2 excluded detectors; and for PSR~B1937+21, 33 of 214 ObsIDs had 1--3 excluded detectors and two ObsIDs had six excluded detectors.  Finally, we filter out photons from short (8~s) stretches of data where the average counts per second are $> 2$. This limit is based on visual inspection of diagnostic plots and is chosen to be permissive: it filters out photons from time periods affected by prominent background, which is typically in the polar horn regions. We prefer this approach over simply excluding all polar horn data since that discards a substantial amount of usable data.

\begin{figure}
\begin{center}
\includegraphics[width=0.6\textwidth]{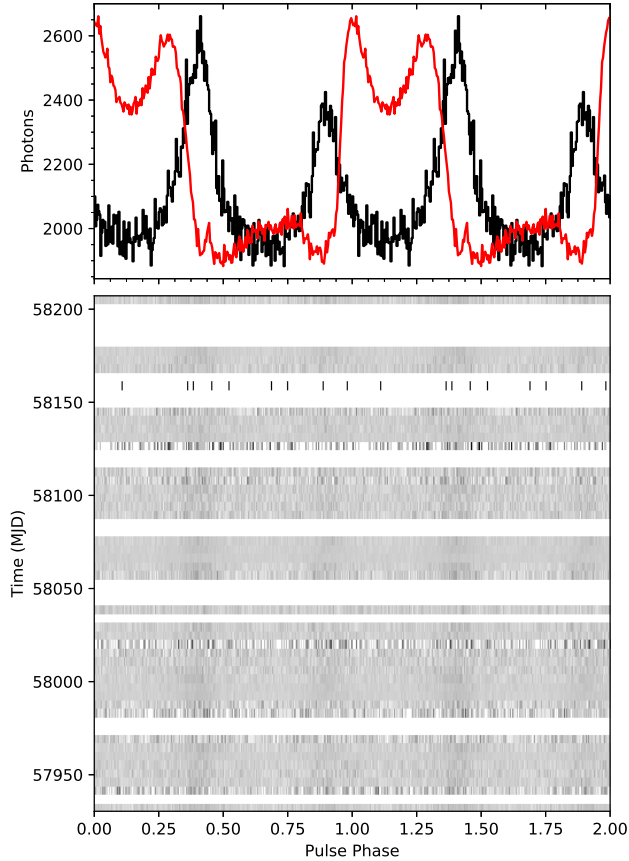}
\caption{Top: The average pulse profile of PSR~J0218+4232 from \textit{NICER} data 
(0.80--6.20~keV) is shown in black. It is phase-aligned with an average 1,484~MHz pulse profile from Nan\c{c}ay Radio Telescope data, shown in red, which has been scaled to the same height. Bottom: Folded \textit{NICER} photons vs.\ pulse phase and MJD. Gaps in the greyscale correspond to no data taken or data excluded by our selection criteria. Each greyscale row represents the same amount of calendar time but the exposure time per row varies. Rows with more exposure time contain more photons and therefore have smoother color variations. In both panels, two full rotations are shown for clarity, with 256 bins per rotation. A DM of 61.2365(6)~pc~cm$^{-3}$ was used in calculating the delay between the frequency of the radio observations and the effective infinite frequency of \textit{NICER} data. \label{fig-0218-prof}}
\end{center}
\end{figure}

\begin{figure}
\begin{center}
\includegraphics[width=0.6\textwidth]{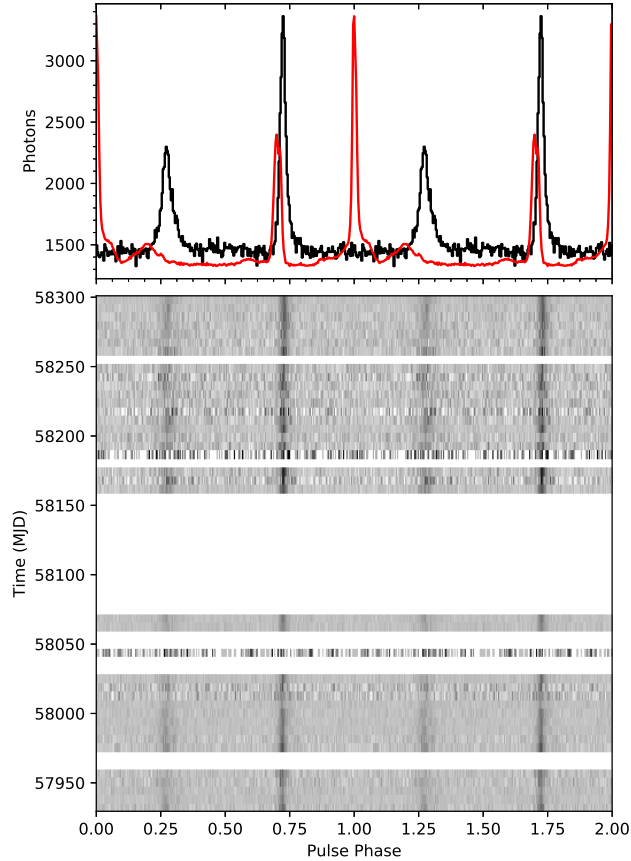}
\caption{Same as Figure~\ref{fig-0218-prof} but for PSR~B1821-24 in the energy range 1.0--5.50~keV, and a DM of 119.8918(7)~pc~cm$^{-3}$. The large gap at MJD~$\sim 58075 - 58160$ corresponds to the pulsar being too close to the Sun to be observed. \label{fig-1821-prof}}
\end{center}
\end{figure}

\begin{figure}
\begin{center}
\includegraphics[width=0.6\textwidth]{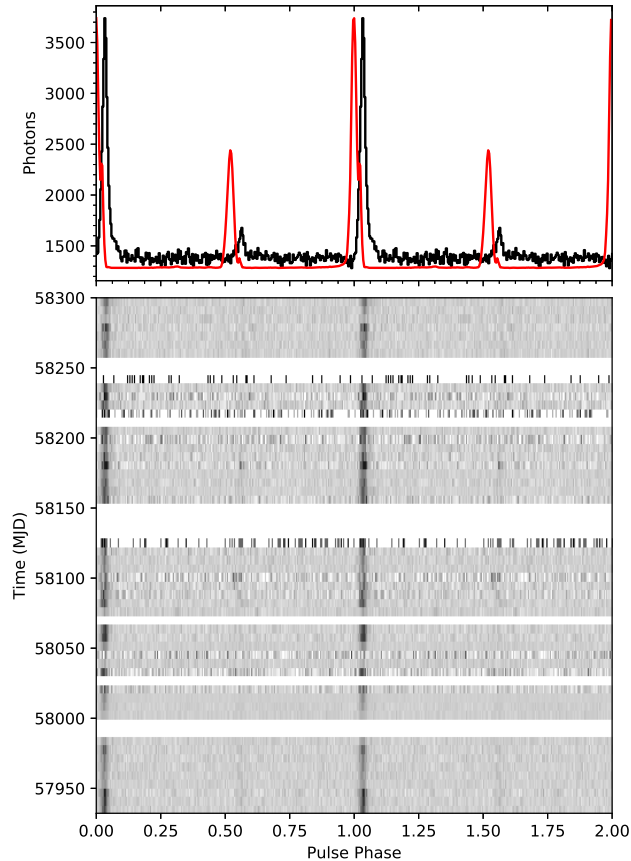}
\caption{Same as Figure~\ref{fig-0218-prof} but for PSR~B1937+21 in the energy range 1.15--5.55~keV, and a DM of 71.01710(9)~pc~cm$^{-3}$. \label{fig-1937-prof}}
\end{center}
\end{figure}

For each pulsar, we derive the optimal photon energy range by computing average pulse profiles and finding the energy bounds that maximize the H-test (\citealt{deJager89}, \citealt{deJager10}), resulting in ranges of 0.8--6.2~keV for PSR~J0218+4232, 1--5.5~keV for PSR~B1821$-$24, and 1.15--5.55~keV for PSR~B1937+21. The profiles are calculated using {\tt PINT}\footnote{\url{https://github.com/nanograv/pint}} with existing phase-connected radio timing solutions for all three pulsars. 
We exclude from further analysis photons outside these energy ranges. The resulting energy-optimized lightcurves are shown in Figures~\ref{fig-0218-prof}, \ref{fig-1821-prof}, and \ref{fig-1937-prof}. The alignment between our X-ray and radio pulse profiles for the three pulsars is consistent with the literature and provides a sanity check for photon phase assignment (\citealt{Knight06}, \citealt{Johnston13}, \citealt{Cusumano04} for PSR~J0218+4232, PSR~B1821$-$21, and PSR~B1937+21, respectively). The figures also show phaseograms of the three pulsars in greyscale, where gaps correspond to no data taken (due to scheduling priorities; in the case of PSR~B1821$-$24, the longest gap corresponds to the pulsar being too close to the Sun to be observed), or data excluded by our selection criteria (e.g., because of high background on some days). 

Overall, for PSR~J0218+4232 we find 538,953 photons satisfy the selection criteria over 722.4~ks of clean exposure time, for an average of 0.75~counts~s$^{-1}$. For PSR~B1821$-$24, 396,435 photons and 447.5~ks of clean exposure remain after selection, with an average of 0.89~counts~s$^{-1}$. For PSR~B1937+21, we get 371,196 photons, 732.0~ks of clean exposure, and 0.51~counts~s$^{-1}$ on average. Note that all of these count rates are total rates including source and background.

\section{The Timing Procedure}\label{sec-timing}

To compare the timing precision and residuals of our three pulsars in X-rays and radio waves we need pulse times of arrival (TOAs) in both energy bands spanning at least several months and overlapping in time. { Table~\ref{tab-setups} lists the radio observing setups used for our three pulsars.} 

For PSR~B1937+21, we use TOAs from the 11-year NANOGrav data set\footnote{ The 11-year NANOGrav TOAs and timing solution for PSR~B1937+21 are available at {\tt http://data.nanograv.org}} obtained from observations with the Arecibo and Green Bank telescopes \citep{11yr}, extended through 2018 January with additional TOAs calculated using the same data-reduction procedures and templates as the 11-year data set. We also use TOAs from the Nan\c{c}ay decimetric Radio Telescope (NRT). Our combined radio TOA set for this pulsar covers $\sim$14\,yr (2004 October 14 -- 2018 January 24). 

{ GBT observations before 2011 used the Green Bank Astronomical Signal Processor (GASP; \citealt{Demorest07}). Arecibo observations before 2012 used the almost identical  Astronomical Signal Processor (ASP). Subsequent observations were recorded using two newer, also nearly identical backends at both observatories: the Green Bank Ultimate Pulsar Processing Instrument (GUPPI; \citealt{DuPlain08}, \citealt{Ford10}) and the Puerto Rican Ultimate Pulsar Processing Instrument (PUPPI). Each backend digitized the baseband voltage signal from the receiver, performed Fourier transforms to convert the time-domain voltage signal into a channelized frequency spectrum, coherently dedispersed the data in all channels in real time, and recorded folded pulse profiles for all channels based on an existing timing solution. ASP and GASP used 4~MHz channels and recorded a folded pulse profile every 60~s. PUPPI and GUPPI used 1.56~MHz channels and recorded a folded profile every 10~s. The folded pulse profiles are summed in time and frequency, and pulse arrival times are extracted for four subbands by comparing the sum to a template using the Fourier-domain technique of \citep{Taylor92}.}  


For PSR~B1821$-$24, we use radio TOAs from observations with the NRT and Parkes telescopes. This combined TOA set spans $\sim$7\,yr (2011 August 31 -- 2018 January 30). { NRT observations were made with the Nan\c{c}ay Ultimate Pulsar Processing Instrument (NUPPI), which is a clone of GUPPI/PUPPI. The NUPPI setup used 4~MHz channels and recorded a folded pulse profile every 10~s. Parkes observations used both the 20\,cm multi-beam receiver and the upper frequency band of the co-axial 10cm/50cm receiver and were recorded with the fourth generation of the Parkes Digital Filterbank System (PDFB4), which measures the four Stokes parameters via polyphase transforms performed on Field-Programmable Gate Array (FPGA) processors (\citealt{Manchester13}, \citealt{Ferris04}). Data are recorded in 1024 channels and recorded to disk every 30s, and observations are preceded by measurements of a pulsed noise diode for complex gain (polarization) calibration.}
For PSR~J0218+4232, we use NRT (NUPPI) TOAs that also span $\sim$7\,yr (2011 August 28 -- 2018 February 1). { The NRT and Parkes TOAs we use in this work will be made publicly available in future European Pulsar Timing Array (EPTA) and Parkes Pulsar Timing Array (PPTA) data releases. } Our \textit{NICER} data for the three pulsars cover the first year of the mission (2017 June 23 -- 2018 June 30). 

\begin{deluxetable}{lccccc}
\tablecaption{ Observing setups for the radio TOAs used in this paper. \label{tab-setups}}
\tablewidth{0pt}
\tablehead{
\colhead{Pulsar} & 
\colhead{MJD range} &
\colhead{Telescope} &
\colhead{$f_{\rm center}$} &
\colhead{Backend} &
\colhead{Bandwidth} \\
 & & & (GHz) & & (MHz)
}
\startdata
PSR~J0218+4232 & 55801--58150 & NRT & 1.4 & NUPPI & 512 \\
\hline
PSR~B1821-24 & 55804--57956 & NRT & 1.4 & NUPPI & 512 \\
             & 55859--57578 & NRT & 2.1 & NUPPI & 512 \\
             & 56810--57774 & NRT & 2.5 & NUPPI & 512 \\
             & 57929--58148 & Parkes & 1.4 & PDFB4 & 256 \\
             & 57929--58148 & Parkes & 3.0 & PDFB4 & 1024 \\
\hline
PSR~B1937+21 & 53420--55974 & Arecibo & 1.4 & ASP & 64 \\
             & 53344--55968 & Arecibo & 2.1 & ASP & 64 \\
             & 56020--58312 & Arecibo & 1.4 & PUPPI & 603 \\
             & 56020--58312 & Arecibo & 2.1 & PUPPI & 460 \\
             & 53275--55243 & GBT & 0.8 & GASP & 64 \\
             & 53267--55390 & GBT & 1.4  & GASP & 48 \\
             & 55278--58302 & GBT & 0.8 & GUPPI & 186 \\
             & 55275--58301 & GBT & 1.4  & GUPPI & 642 \\
             & 55800--58367 & NRT & 1.4 & NUPPI & 512 \\
             & 55804--57574 & NRT & 2.1 & NUPPI & 512 \\
             & 56510--58301 & NRT & 2.5 & NUPPI & 512 \\
\enddata
\end{deluxetable}

\begin{deluxetable}{lccc}
\tablecaption{Two-Gaussian noiseless templates used to extract TOAs from \textit{NICER} data. The phase and FWHM are in fractional phase units. The amplitude of each component is the fraction of the total counts accounted for by that component.\label{tab-gaussians}}
\tablewidth{0pt}
\tablehead{
\colhead{Pulsar} & 
\colhead{Phase} &
\colhead{Amplitude} &
\colhead{FWHM} 
}
\startdata
PSR~J0218+4232 & 0.0 & 0.02252 & 0.11956 \\
               & 0.50063 & 0.03845 & 0.13552 \\
PSR~B1821$-$24   & 0.0 & 0.03171 & 0.02729 \\
               & 0.55037 & 0.02484 & 0.05216 \\
PSR~B1937+21   & 0.0 & 0.03810 & 0.02514 \\
               & 0.53059 & 0.00275 & 0.01177 \\              
\enddata
\end{deluxetable}

Pulses differ in their shape both between energy bands and from one pulse to another within each energy band.
However, the average radio and X-ray pulse shapes of our three pulsars are very stable with time. This stability means that for each observing setup we can construct a noiseless template approximating the average pulse profile that is then used to extract TOAs from data taken with that setup. Our three pulsars have X-ray pulse profiles with a main pulse and interpulse $\sim$180$\degrees$ apart in phase. TOA quality is maximized if we can construct a pulse shape template that fits multiple narrow resolved features. When we extract TOAs, we use a pulse shape template constructed by fitting two Gaussians to the average pulse profiles from our data span (the black \textit{NICER} profiles from the top panels of Figures~\ref{fig-0218-prof}--\ref{fig-1937-prof}).
Compared to using a single-Gaussian template, this also minimizes the chance that some TOAs will be calculated with respect to the main pulse while others will be calculated with respect to the interpulse.  
Table~\ref{tab-gaussians} shows the parameters of our noiseless templates for all three pulsars. In these templates, one of the peaks is centered at zero phase. In contrast, Figures~\ref{fig-0218-prof}--\ref{fig-1937-prof} show the radio and X-ray folded pulse profiles aligned in absolute phase, which is determined based on the radio timing solution of each pulsar. Hence the radio peak of each pulsar is at zero phase in the combined, phase-aligned plots. 

While radio data are recorded as regularly sampled time series, in X-rays we have sparse photons whose individual detection times follow Poisson statistics. Moreover, only a small percentage of the photons detected during a pulsar observation come from the pulsar as opposed to background. We use on-board GPS measurements, recorded every 10~s, to interpolate \textit{NICER}'s position and velocity with respect to the geocenter at each photon detection time, which is measured in the spacecraft-topocentric frame. 

In order to extract an X-ray TOA from \textit{NICER} data, we barycenter photon time stamps and assign a phase to each photon using {\tt PINT}'s {\tt photonphase} routine together with the radio timing solution, which is phase-connected over the multi-year time span of the radio TOAs. We then construct a histogram of photon phases for each ObsID similar to the top panels of Figures~\ref{fig-0218-prof}--\ref{fig-1937-prof}, which show such histograms for our full \textit{NICER} data span. We obtain one TOA per ObsID by using a two-Gaussian X-ray template and the maximum likelihood method of \citet{Ray11}. The reference time for each TOA is the photon time stamp closest to the middle of the range of photon time stamps included in the ObsID. \textit{NICER} spacecraft-topocentric TOAs are recorded and used together with the spacecraft's geocentric position and velocity at each TOA. This is analogous to treating ground-based TOAs, where the first transformation is from the observatory's reference frame to the geocenter and uses an ephemeris of the Earth's rotation. { We include our \textit{NICER} spacecraft-topocentric TOAs as supplementary electronic materials.}

In our timing work we use the DE430 JPL ephemeris, Barycentric Dynamical time (TDB) units, and the TT(BIPM2015) clock realization. 

For ObsIDs with total GTIs $\lesssim 100$~s, there are not enough photons for the pulse peak to be significantly detected in the resulting histogram. These TOAs are excluded from the timing analysis. 

\section{TOA Uncertainties}\label{sec-toa-errors}

In order to evaluate the quality of \textit{NICER} TOAs, we compare the actual TOA uncertainties with estimates from theory and simulations. Even though our pulse shape templates consist of two Gaussians, as a first approximation it is useful to evaluate the single-Gaussian case analytically using only the Gaussian fitted to the higher peak. The simplest way to estimate TOA uncertainty is to calculate to what accuracy $\sigma_{\rm T}$ we can localize the centroid of a Gaussian in the presence of background emission as well as statistical noise in the arrival times of source photons. This is given by the ratio of the pulse width to the pulse signal-to-noise ratio. Photon detections by \textit{NICER} follow Poisson statistics, therefore the pulse signal-to-noise in terms of total source and background photon counts within an observation is $N_{\rm src}/\sqrt{N_{\rm src}+N_{\rm bkg}}$. If $P$ is the pulse period, $w$ is the standard deviation of the Gaussian in terms of pulse phase, $T$ is the time span of photons yielding a single TOA, $\alpha$ is the average source photon count per second, and $\beta$ is the average background photon count per second, $\sigma_{\rm T}$ is given by
\be
\sigma_{\rm T} = \left(P w\right) \sqrt{\frac{\alpha+\beta}{\alpha}}\frac{1}{\sqrt{\alpha T}}\label{eq-simple-gaussian}
\ee
(e.g., \citealt{Handbook}).

However, all three pulsars have two-peak pulse profiles which are better approximated as a sum of Gaussians with different amplitudes and standard deviations. Since the rate at which pulsars emit photons is phase-dependent, photon arrival times can be modeled as a non-homogeneous Poisson process whose rate function is the two-Gaussian pulse template $h(\phi)$, where the pulse phase $\phi \in \left[0,1\right)$ and $\int_0^1 h\left(\phi\right) d\phi = 1$. In this case, the lower limit on TOA uncertainty corresponds to the Cramer-Rao Lower Bound\footnote{The Cramer-Rao Lower Bound is a lower bound on the variance of an unbiased estimator of a fixed but unknown parameter.} (CRLB), as shown by \cite{Golshan07} and \cite{Winternitz16}. \citeauthor{Golshan07} derive the CRLB as
\be
\sigma_{\rm T} \geq \frac{P}{\sqrt{I_p T}}{\rm, where}\label{eq-crlb}
\ee
\be
I_p = \int_0^1 \frac{\left[\alpha~\frac{\partial}{\partial \phi} h\left(\phi\right) \right]^2}{\beta + \alpha h\left(\phi \right)} {\rm d}\phi.\label{eq-ip-numerical}
\ee
In the case of pulsars where $w << 1$ and the background is low, this can be simplified by considering only the main pulse peak and setting $\beta = 0$, so that $I_p = \alpha/w^2$.


In addition to these theoretical estimates, we simulated TOAs based on \textit{NICER} instrument design parameters and average source and background photon count rates for each pulsar. { We used inverse transform sampling \citep{Devroye86}, a technique for generating random numbers from a probability density function, to simulate pulsed events.} We did this over a wide variety of integration times taking into account the predicted pulsed and unpulsed count rates and smooth two-Gaussian models of the X-ray pulse profiles, which we used as photon arrival probability density functions as a function of pulse phase. Unpulsed photons were simulated by drawing uniform random deviates over the full range of the pulse phase. Time-integrated TOAs were determined from the simulated events using the unbinned maximum likelihood technique \citep{Ray11}, with the smooth pulse profile model as the template, and TOA errors computed from the second moment of the resulting likelihood function. 

\begin{figure}
\begin{center}
\includegraphics[width=0.75\textwidth]{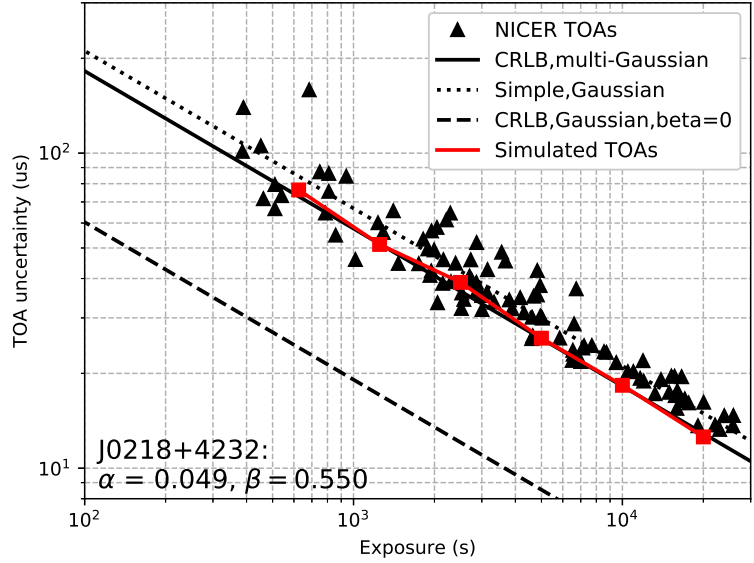}
\caption{\textit{NICER} TOAs vs.\ TOA uncertainty and exposure time for PSR~J0218+4232 (black triangles) are shown together with the simulated TOAs (red) and theoretical estimates described in Section~\ref{sec-toa-errors}. The solid line corresponds to the numerical CRLB result based on the two-Gaussian pulse template (Eqs.~\ref{eq-crlb} and \ref{eq-ip-numerical}). Analytical results for a single Gaussian are shown with a dotted line for a simple estimate based on the ratio of pulse width and signal-to-noise (Eq.~\ref{eq-simple-gaussian}), and with a dashed line for the CRLB assuming zero background. 
\label{fig-0218-crlb}}
\end{center}
\end{figure}

\begin{figure}
\begin{center}
\includegraphics[width=0.75\textwidth]{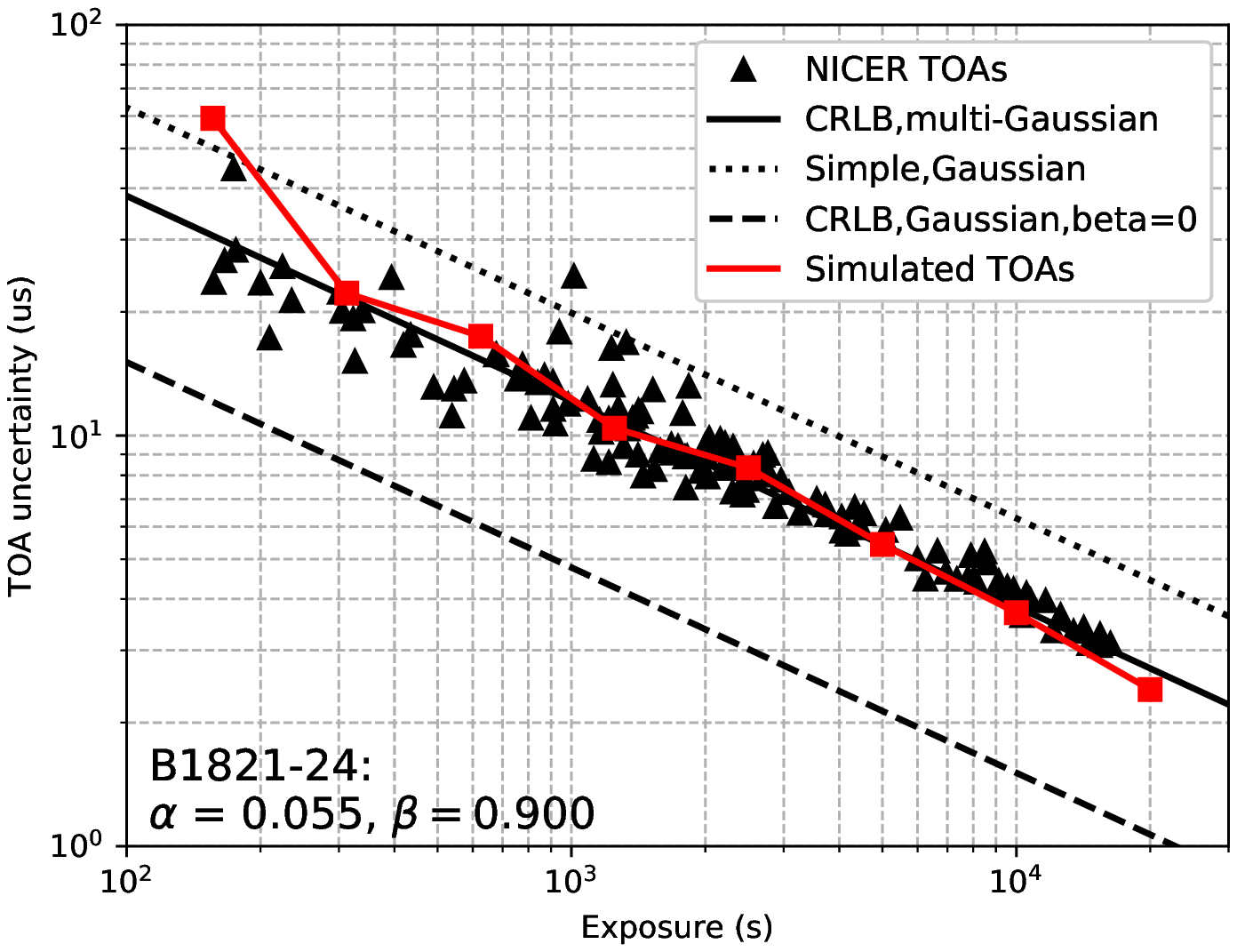}
\caption{\textit{NICER} TOAs vs. TOA uncertainty and exposure time for PSR~B1821$-$24 (black triangles) are shown together with the simulated TOAs (red) and theoretical estimates described in Section~\ref{sec-toa-errors}.\label{fig-1821-crlb}}
\end{center}
\end{figure}

\begin{figure}
\begin{center}
\includegraphics[width=0.75\textwidth]{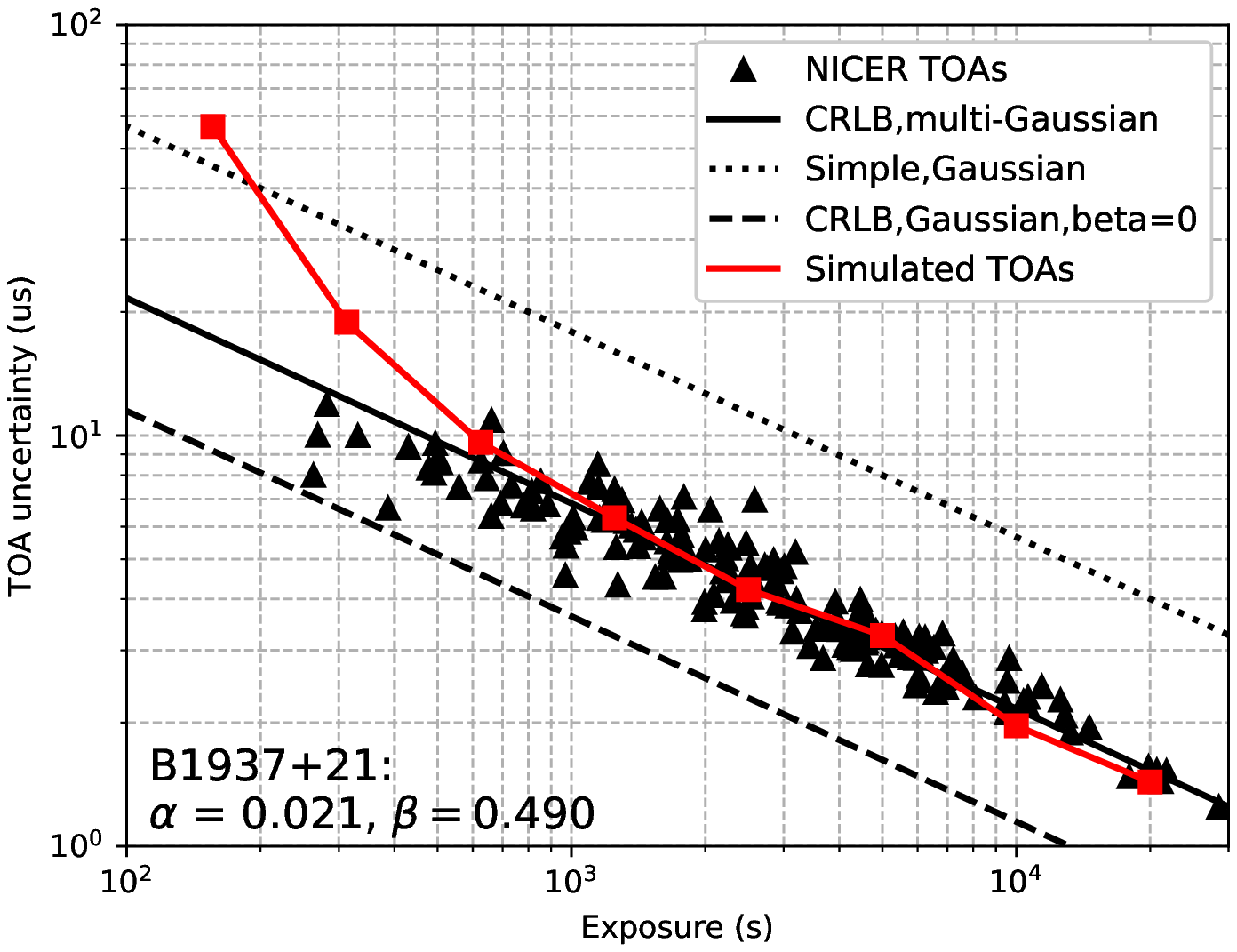}
\caption{\textit{NICER} TOAs vs. TOA uncertainty and exposure time for PSR~B1937+21 (black triangles) are shown together with the simulated TOAs (red) and theoretical estimates described in Section~\ref{sec-toa-errors}.\label{fig-1937-crlb}}
\end{center}
\end{figure}

Figures~\ref{fig-0218-crlb}--\ref{fig-1937-crlb} show the uncertainties of actual \textit{NICER} TOAs (where each TOA is based on data from one ObsID) and simulations. We also show estimates based on two simplified, one-Gaussian cases: Equation~\ref{eq-simple-gaussian}, and Equation~\ref{eq-crlb} evaluated assuming $\beta=0$. 
Finally, the most realistic theoretical estimate of TOA uncertainties is from the numerically evaluated CRLB using the two-Gaussian pulse template for each pulsar and Equation~\ref{eq-crlb}. We find that our TOAs are consistent with the CRLB numerical lower limit as well as with simulations. 

\section{Radio and X-ray Timing Residuals}\label{sec-residuals}

Pulsar timing models include a parameter called dispersion measure (DM), which is the integrated column density of ionized gas along the line of sight to the pulsar. The corresponding dispersion delay reflected in TOAs is $\propto {\rm DM}~\nu^{-2}$ \citep{Handbook}. When TOAs span many years, as is the case for our radio data, the observed DM slowly and stochastically varies with time due to the changing line of sight to the pulsar and turbulence in the ISM. Any unmodelled variations manifest as red noise (i.e., noise whose spectral density is higher at lower frequencies than at high frequencies) in radio timing residuals on a time scale of months to years. X-ray TOAs are effectively at infinite electromagnetic frequency compared to radio TOAs, which means that they are immune to dispersive, diffractive, and refractive (e.g., \citealt{Shannon17}) propagation effects due to ionized gas. 

Along with any timing noise intrinsic to a pulsar, DM variations are a limiting factor for radio pulsar timing precision. DM variations can be modeled either by including time-derivatives of DM as additional parameters in the timing solution or by fitting offsets (DMX) to the best-fit DM at a reference epoch for TOA subsets that span a shorter period of time. DM derivatives tend to be highly covariant, unlike DMX offsets. We follow NANOGrav methods and opt for the DMX approach \citep{11yr}, as we can more easily extend the timing solution in the future as the data span grows. 

For each pulsar, we begin with only radio TOAs and the corresponding best-fit timing solution that does not include DM derivatives or rotational frequency derivatives higher than first order. In the case of PSR~B1937+21, this is the solution from the 11-year NANOGrav data release\footnote{\url{https://data.nanograv.org}}, which includes DMX fits for 6-day intervals. Systematics in the NANOGrav and NRT TOA sets on PSR~B1937+21 make it difficult to combine them in an unbiased manner for the purpose of measuring pulsar stability, and we treat them separately. Because both TOA sets are densely sampled and $>6$\,yr long, two independent measurements of the stability of PSR~B1937+21 are available. 

We follow the procedure described in \cite{11yr} to generate new DMX intervals that cover the span of our radio TOAs for all three pulsars. We tailor the DMX interval length to the observation cadence for each pulsar as well as how often multi-frequency radio observations occur. In the absence of multi-frequency observations DMX offsets can be fitted using TOAs from several frequency subbands of the same observation. However, using TOAs from at least two different observing frequency bands that cover a wider range in frequency allows for a better DMX fit. The resulting DMX interval lengths are 6 days for PSR~B1937+21, 15 days for PSR~B1821$-$24, and 30 days for PSR~J0218+4232. We disable fitting for all parameters except the DMX offsets and insert JUMPs\footnote{A JUMP is a fitted parameter that accounts for phase offsets between data sets.} every three days. The JUMPs effectively bracket each observing epoch, while allowing for epochs that may span MJDs or coordinated (usually, multi-frequency) observations on adjacent days. Typical, non-coordinated observations are a week or more apart. While DMX parameters will preferentially absorb chromatic red noise caused by DM variations, JUMPs will preferentially absorb achromatic red noise caused by intrinsic rotational instabilities. This is equivalent to the approach taken by \cite{Taylor91} and \cite{Kaspi94}. However, the JUMPs and DMX offsets are still covariant. While the best fit will minimize the root-mean-square of the timing residuals, the resulting noise term separation cannot be verified to correspond to the real red noise contributions of DM variations vs.\ intrinsic rotational instabilities by using only radio data. 

\begin{figure}
\begin{center}
\includegraphics[clip,trim={0 0 0 30},width=0.7\textwidth]{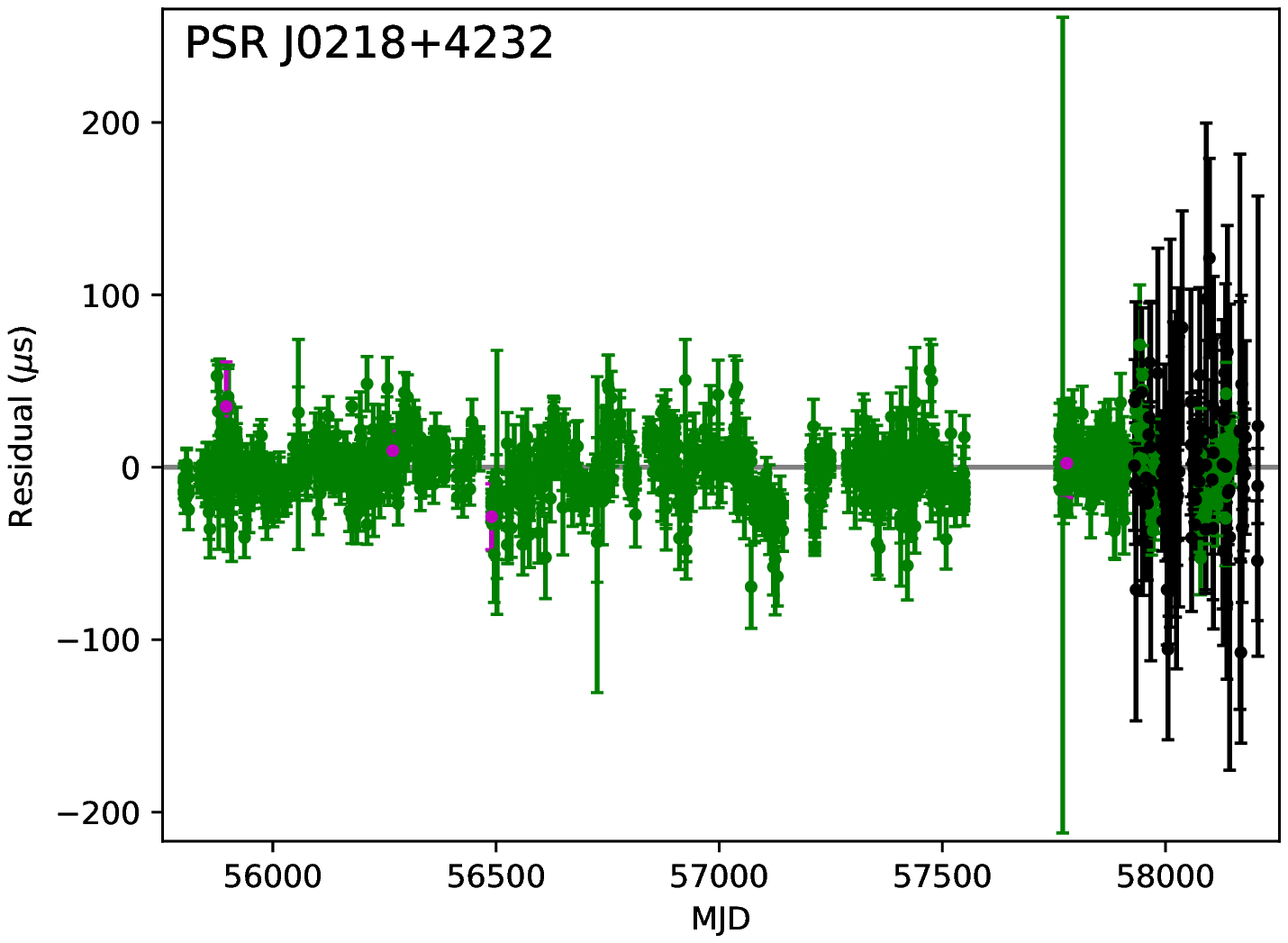}\vspace{-0.5cm}
\includegraphics[clip,trim={0 0 0 30},width=0.7\textwidth]{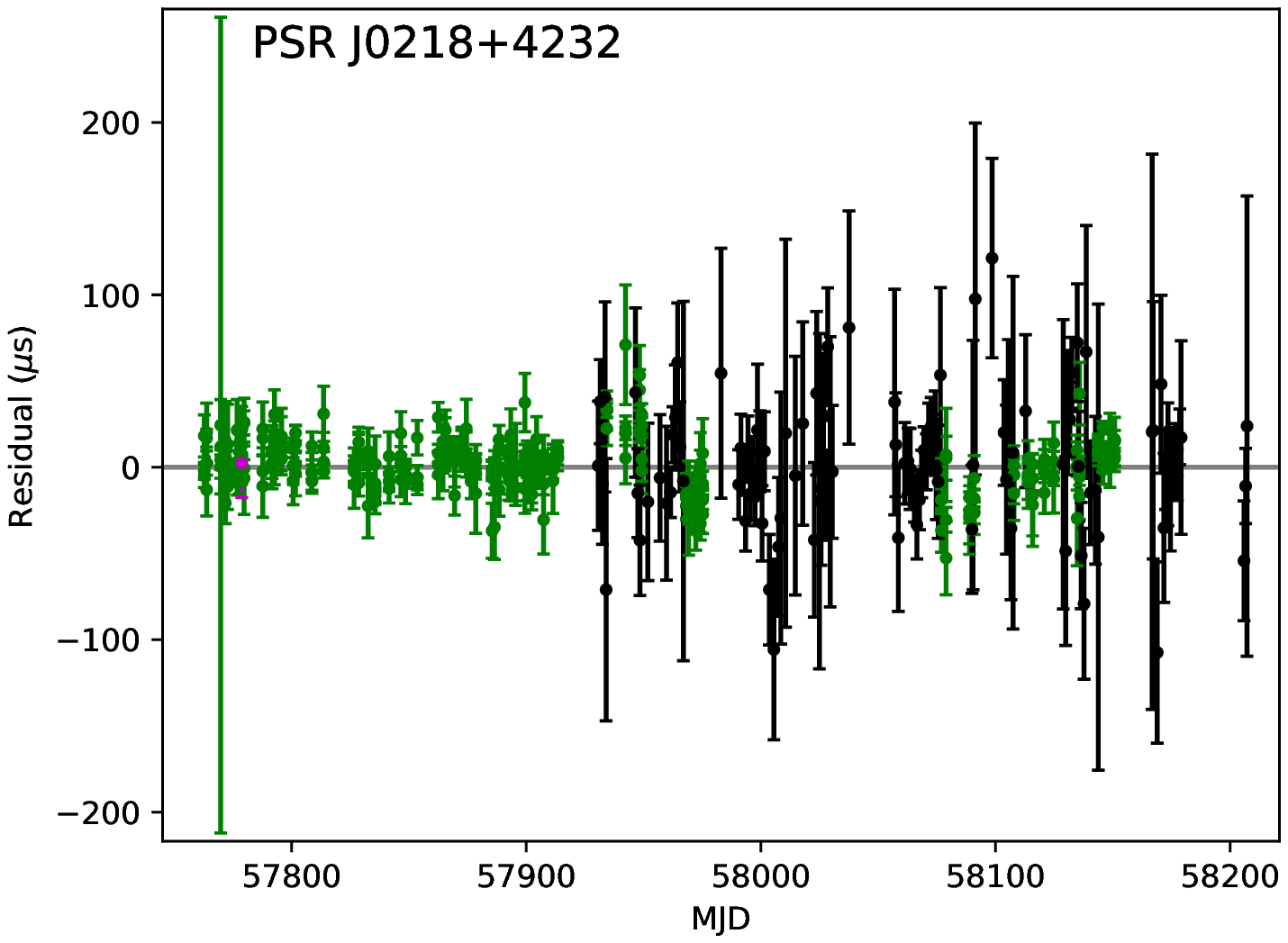}
\caption{PSR~J0218+4232 timing residuals, from fits to the radio data only, for the full radio and X-ray data spans (top) and a zoomed-in range around the \textit{NICER} data (bottom). \textit{NICER} residuals are shown in black; NRT residuals are shown in green (1.0--1.7~GHz) and magenta (1.7--2.7~GHz).\label{fig-0218-res}}
\end{center}
\end{figure}

\begin{figure}
\begin{center}
\includegraphics[clip,trim={0 0 0 35},width=0.7\textwidth]{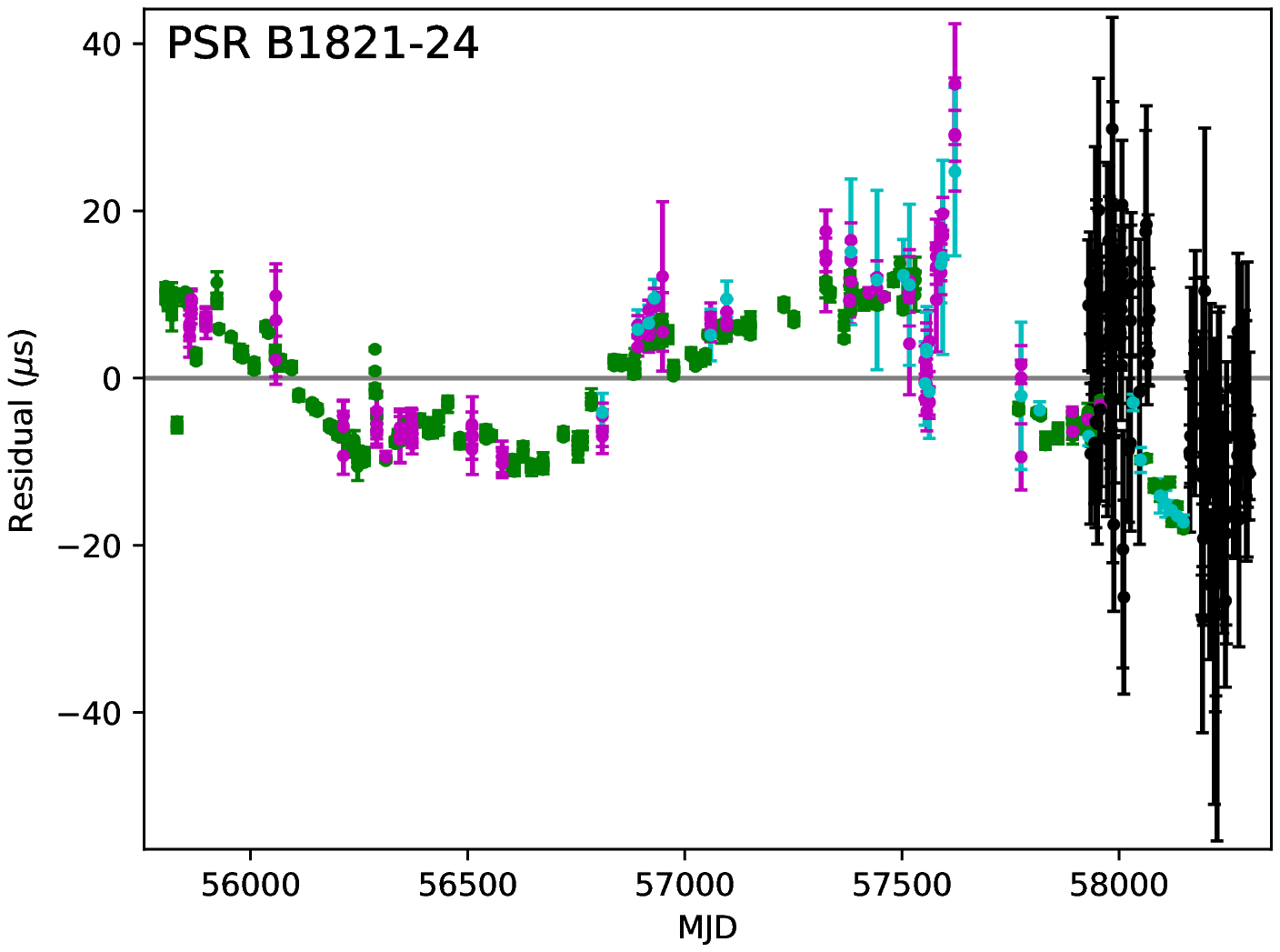}\vspace{-0.5cm}
\includegraphics[clip,trim={0 0 0 35},width=0.7\textwidth]{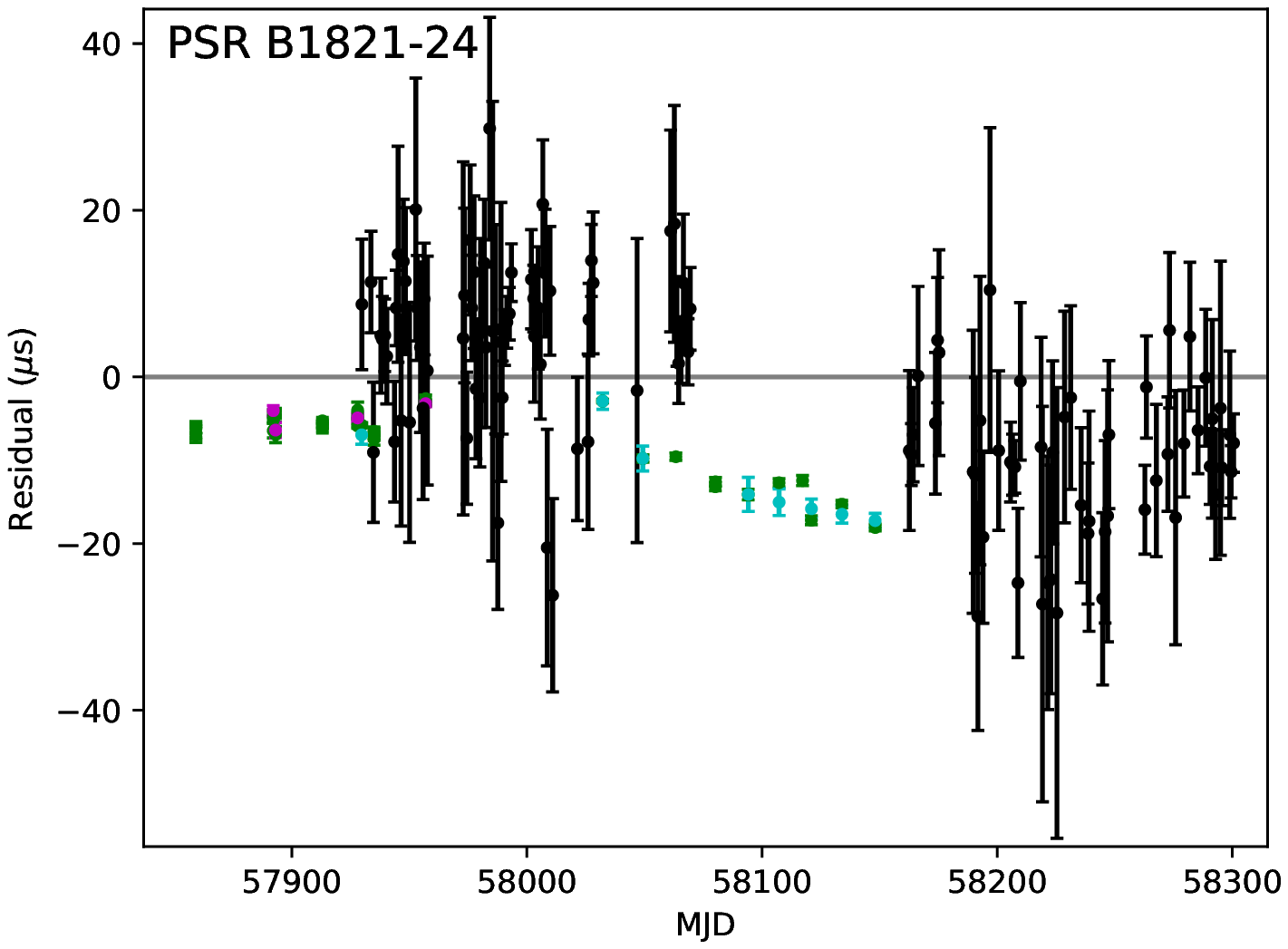}
\caption{PSR~B1821$-$24 timing residuals, from fits to the radio data only, for the full radio and X-ray data spans (top) and a zoomed-in range around the \textit{NICER} data (bottom). \textit{NICER} residuals are shown in black. NRT and Parkes residuals are shown in green (1.0--1.7~GHz), magenta (1.7--2.7~GHz), and cyan ($> 2.7$~GHz). \label{fig-1821-res}}
\end{center}
\end{figure}

\begin{figure}
\begin{center}
\includegraphics[clip,trim={0 0 0 35},width=0.7\textwidth]{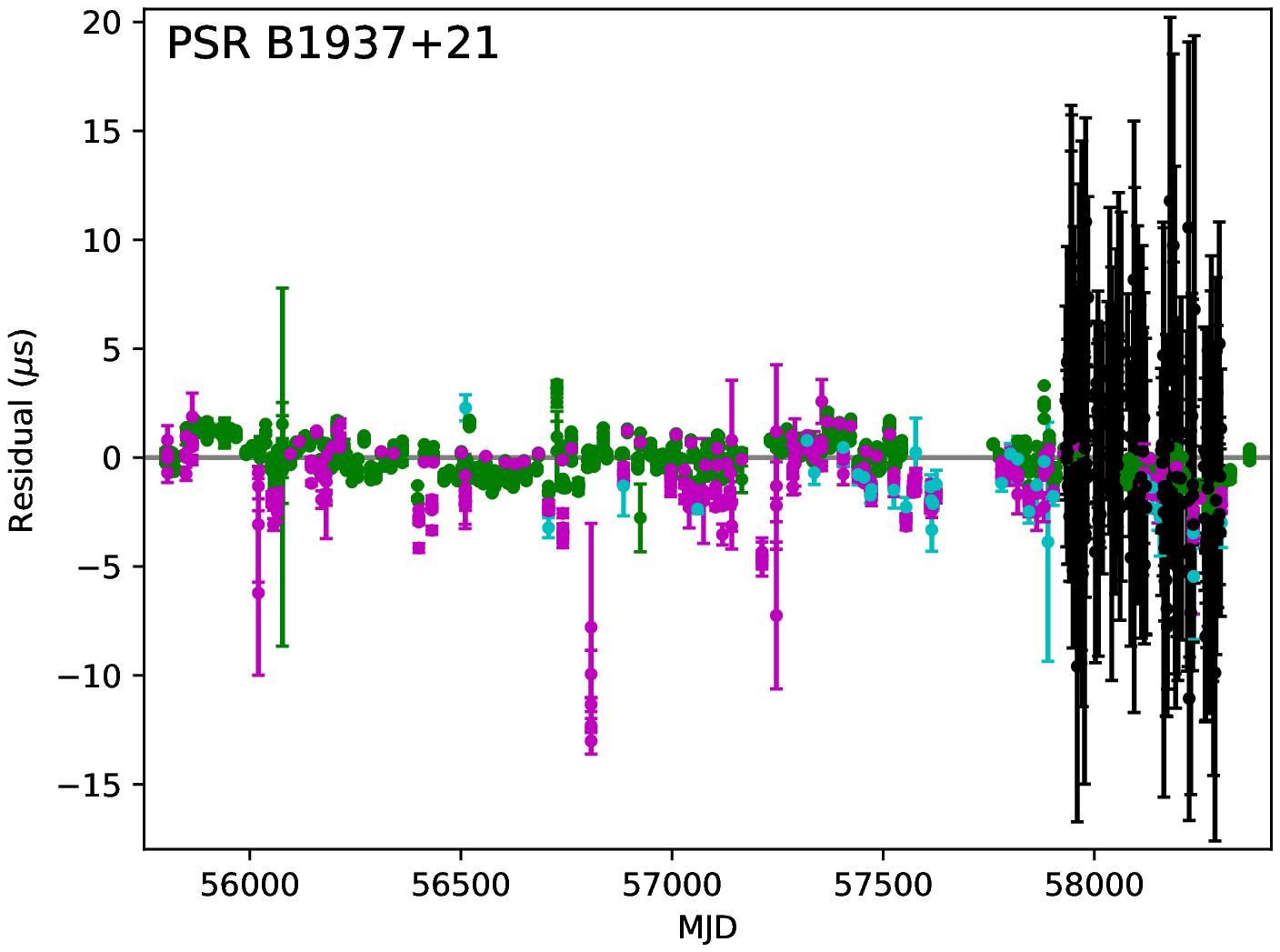}\vspace{-0.5cm}
\includegraphics[clip,trim={0 0 0 35},width=0.7\textwidth]{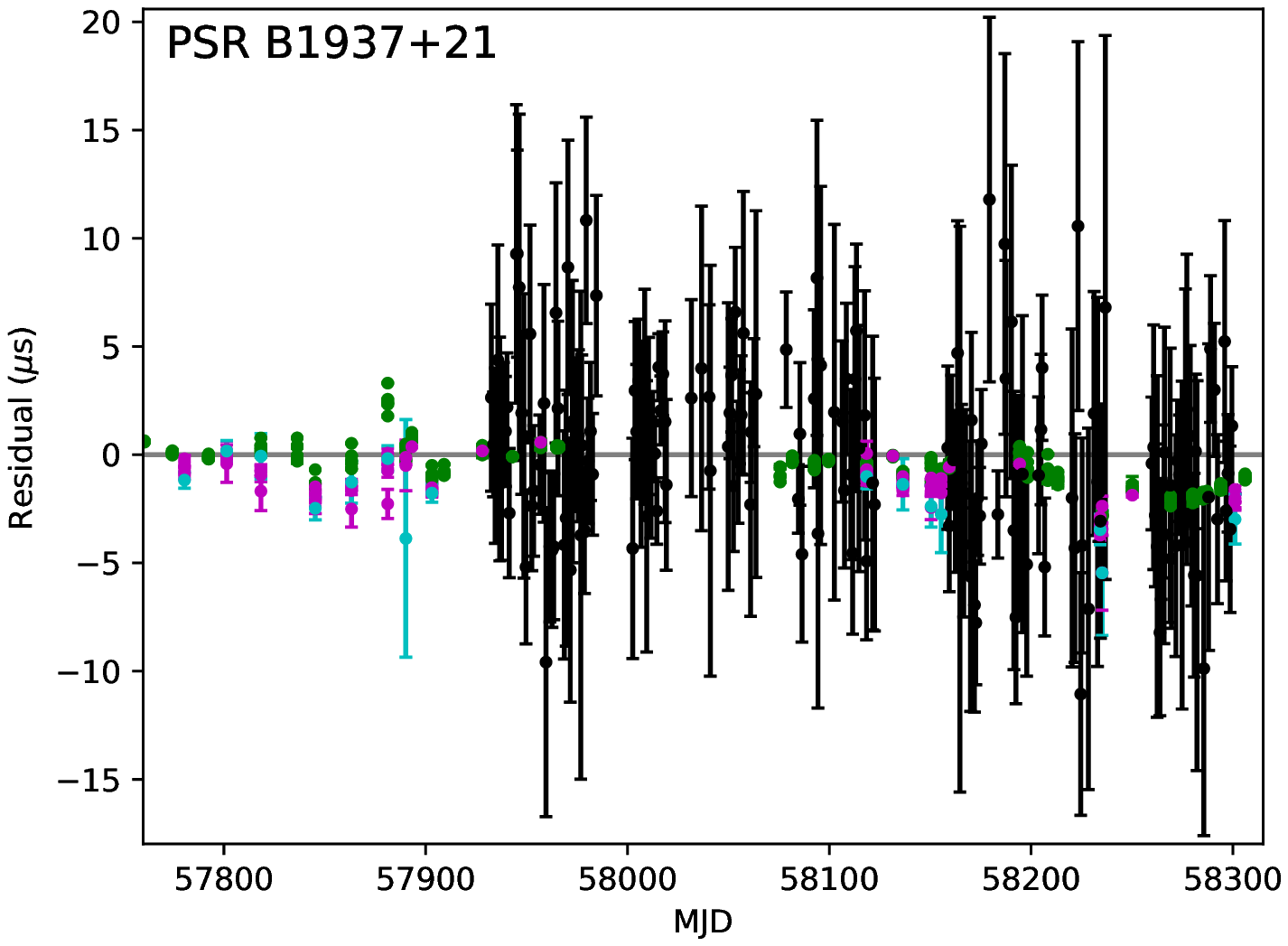}
\caption{PSR~B1937$+$21 timing residuals, from fits to the radio data only, for the full NRT and X-ray data spans (top) and a zoomed-in range around the \textit{NICER} data (bottom). \textit{NICER} residuals are shown in black. NRT residuals are shown in green (1.0--1.7~GHz), magenta (1.7--2.7~GHz), and cyan ($> 2.7$~GHz). \label{fig-1937-res-nancay}}
\end{center}
\end{figure}

\begin{figure}
\begin{center}
\includegraphics[clip,trim={0 0 0 35},width=0.7\textwidth]{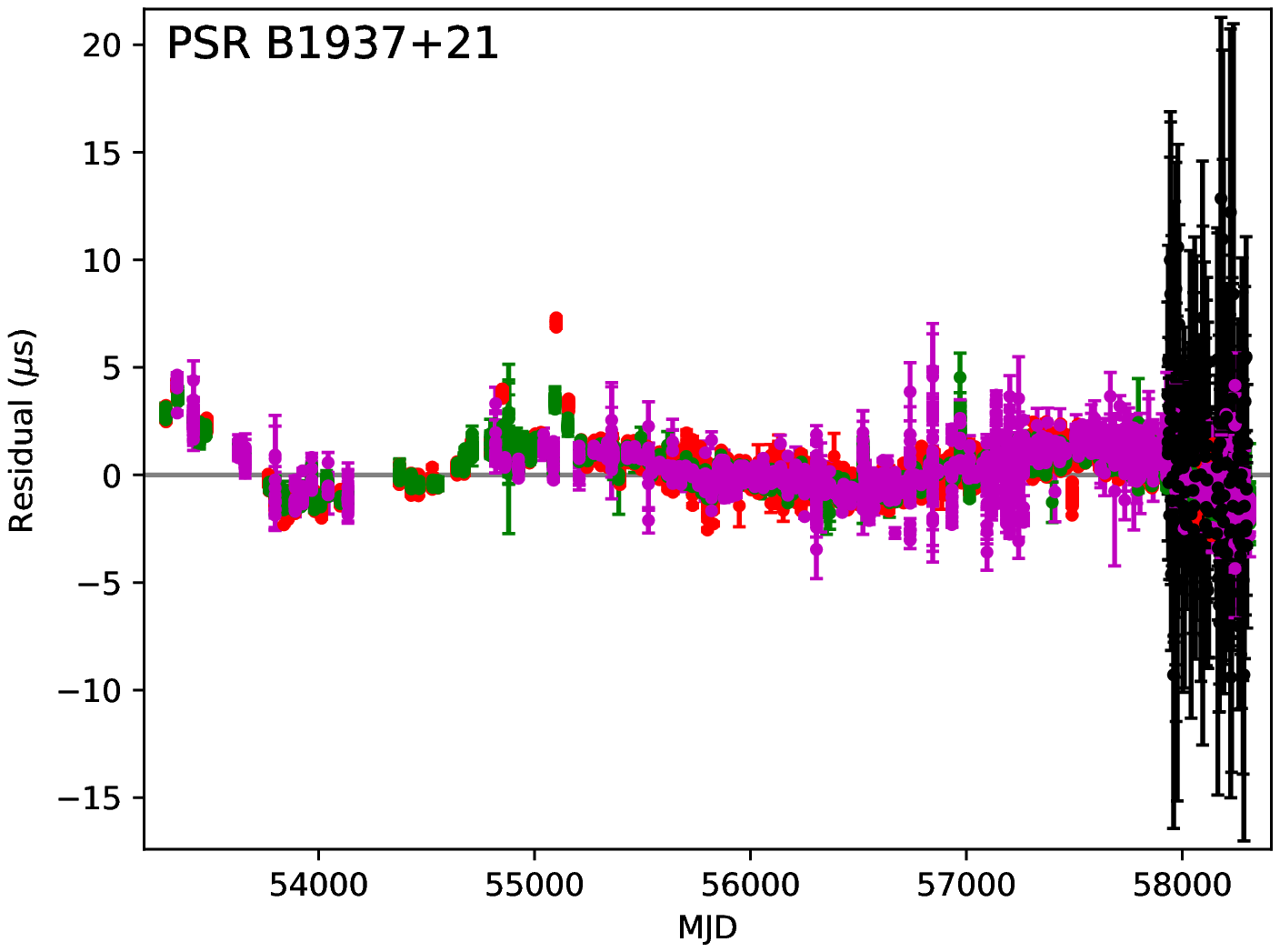}\vspace{-0.5cm}
\includegraphics[clip,trim={0 0 0 35},width=0.7\textwidth]{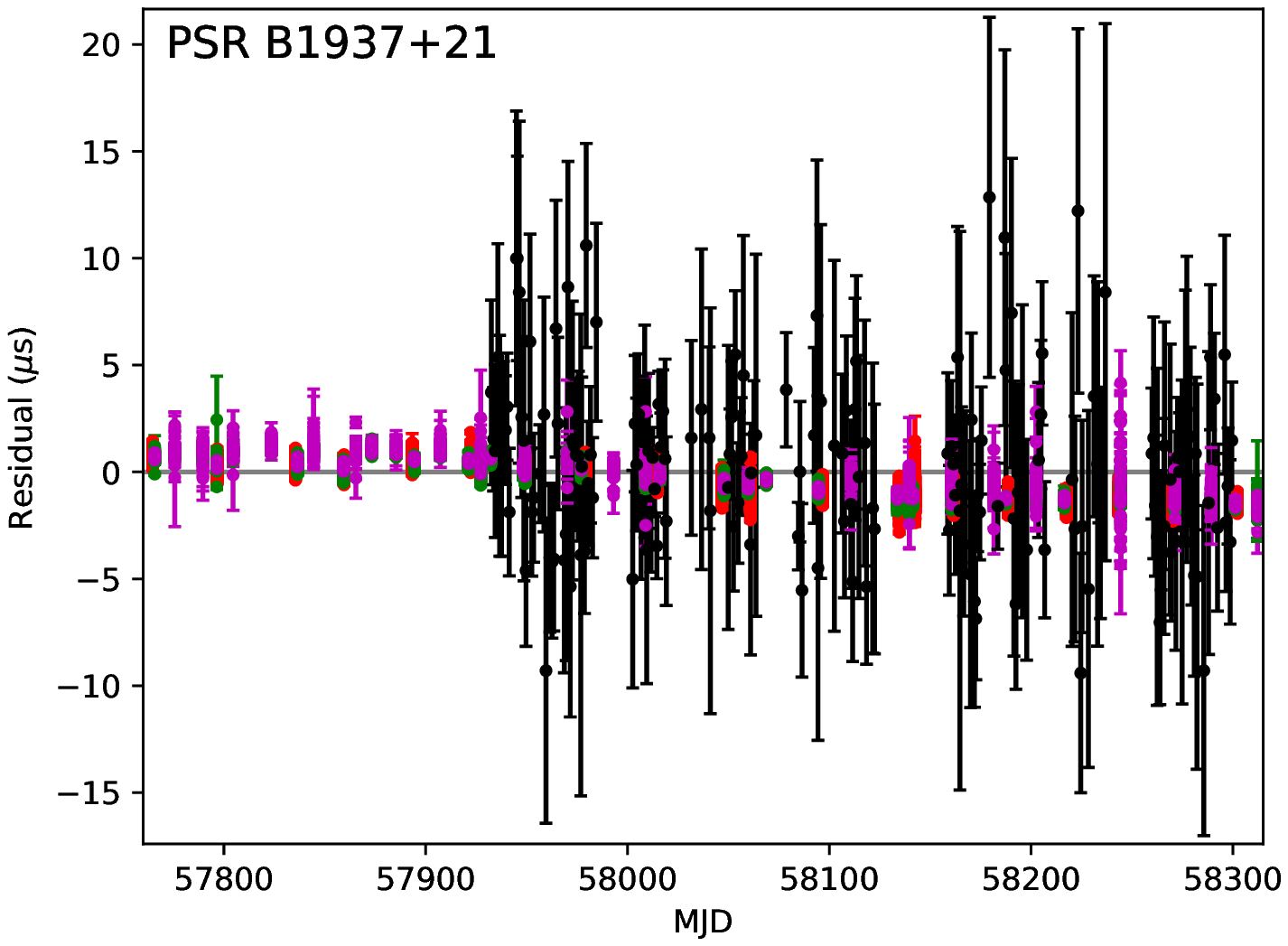}
\caption{PSR~B1937$+$21 timing residuals, from fits to the radio data only, for the full NANOGrav and X-ray data spans (top) and a zoomed-in range around the \textit{NICER} data (bottom). \textit{NICER} residuals are shown in black. NANOGrav residuals are shown in red (0.5--1.0~GHz), green (1.0--1.7~GHz), and magenta (1.7--2.7~GHz). \label{fig-1937-res-nanograv}}
\end{center}
\end{figure}

Next we use Tempo\footnote{\url{http://tempo.sourceforge.net/}} to obtain a fit for the DMX offsets; disable fitting for the DMX offsets and enable it for all other parameters excluding DM; and remove the three-day JUMPs. After we refit again, it is the fitted rotational, astrometric, and binary parameters that absorb the red noise contribution previously absorbed by the JUMPs. Figures~\ref{fig-0218-res}--\ref{fig-1937-res-nanograv} show the effects of this on the residuals in the form of smoothly varying deviations from zero residual on a time scale of months to years. { We include these final radio timing solutions as supplementary electronic materials.}


In order to include \textit{NICER} TOAs in our updated timing solution, we fit a JUMP to account for the phase difference between the radio and X-ray pulse shown in Figures~\ref{fig-0218-prof}, \ref{fig-1821-prof}, and \ref{fig-1937-prof}. Figures~\ref{fig-0218-res}--\ref{fig-1937-res-nanograv} show \textit{NICER} and radio timing residuals with respect to the final radio timing solution from the fitting steps described above. The top panel of each figure shows the entire data span, and the bottom panel shows a zoomed area of overlap between \textit{NICER} and radio TOAs. \textit{NICER} TOAs are shown in black, and radio TOAs are colored according to the observing frequency. The slight remaining variations and inconsistencies between radio residuals at different observing frequencies, even after DMX offset fitting, reflect the difficulty in removing propagation effects, which is a limitation on radio timing precision. 

\section{Rotational Stability Estimates}\label{sec-sigmaz}

When characterizing clock stability, the statistic of choice is typically the Allan variance, $\sigma^2_y$, which depends on second differences between clock frequency offset measurements:
\be
\sigma_y^2 = \left< \frac{1}{2} \left(\bar{y}_n - \bar{y}_{n-1} \right)^2\right>,
\ee
where $\bar{y}_n$ is the average fractional clock frequency offset during the $n$-th measurement interval of a certain length, and the angle brackets denote an average over all intervals of the same length. 
The Allan variance is designed to quantify instability in clocks that operate at constant frequency. However, because of pulsars' continuous energy loss, manifested as an observed spin period derivative, they act as clocks with linearly varying frequency at an {\it a priori} unknown drift rate. Therefore the ``clock noise'' must be quantified in a different way.

When we obtain a best-fit timing solution for a pulsar over a span of at least one year, first-order deviations from the unknown actual pulse period are modeled and removed via fitting the pulse period, period derivative, and pulsar position on the sky. 
For multi-year high-precision MSP timing solutions, such as the radio fits used in this paper, we can also remove the effects of proper motion and parallax. We are interested in characterizing the remaining timing residual perturbations, whose lowest-order term is cubic. It is caused by intrinsic timing noise due, e.g., to rotational instabilities and, in the case of radio timing, imperfectly modeled propagation effects \citep{Cordes13}. Timing noise tends to have a ``red'' power spectrum, and a better measure for pulsar stability would be one that is more sensitive to red noise. \cite{Matsakis97} introduce such a measure based on third-order variations in the timing residuals: $\sigma_z$. 

For an interval of length $\tau$ starting at time $t_0$, we can fit a cubic polynomial to timing residuals in that interval,
\be
X(t) = c_0+c_1(t-t_0)+c_2(t-t_{0})^2 +c_3(t-t_0)^3,
\ee
where $X(t)$ minimizes the sum of $\left[(x_i - X(t_i))/\sigma_i \right]^2$ over all TOAs $t_i$ with uncertainties $\sigma_i$ and residuals $x_i$. Then 
\be
\sigma_z \equiv \frac{\tau^2}{2\sqrt{5}}\left<c_3^2\right>^{1/2}\label{eqn-sigmaz},
\ee
where angle brackets denote the weighted average over the third-order coefficients of the best-fit polynomials of all non-overlapping intervals of length $\tau$ within the set of timing residuals. In our analysis we follow the recipe for calculating $\sigma_z$ in \cite{Matsakis97} and compute separate $\sigma_z$ values for radio and \textit{NICER} TOAs for each of our three pulsars. 

\begin{figure}
\begin{center}
\includegraphics[width=0.75\textwidth]{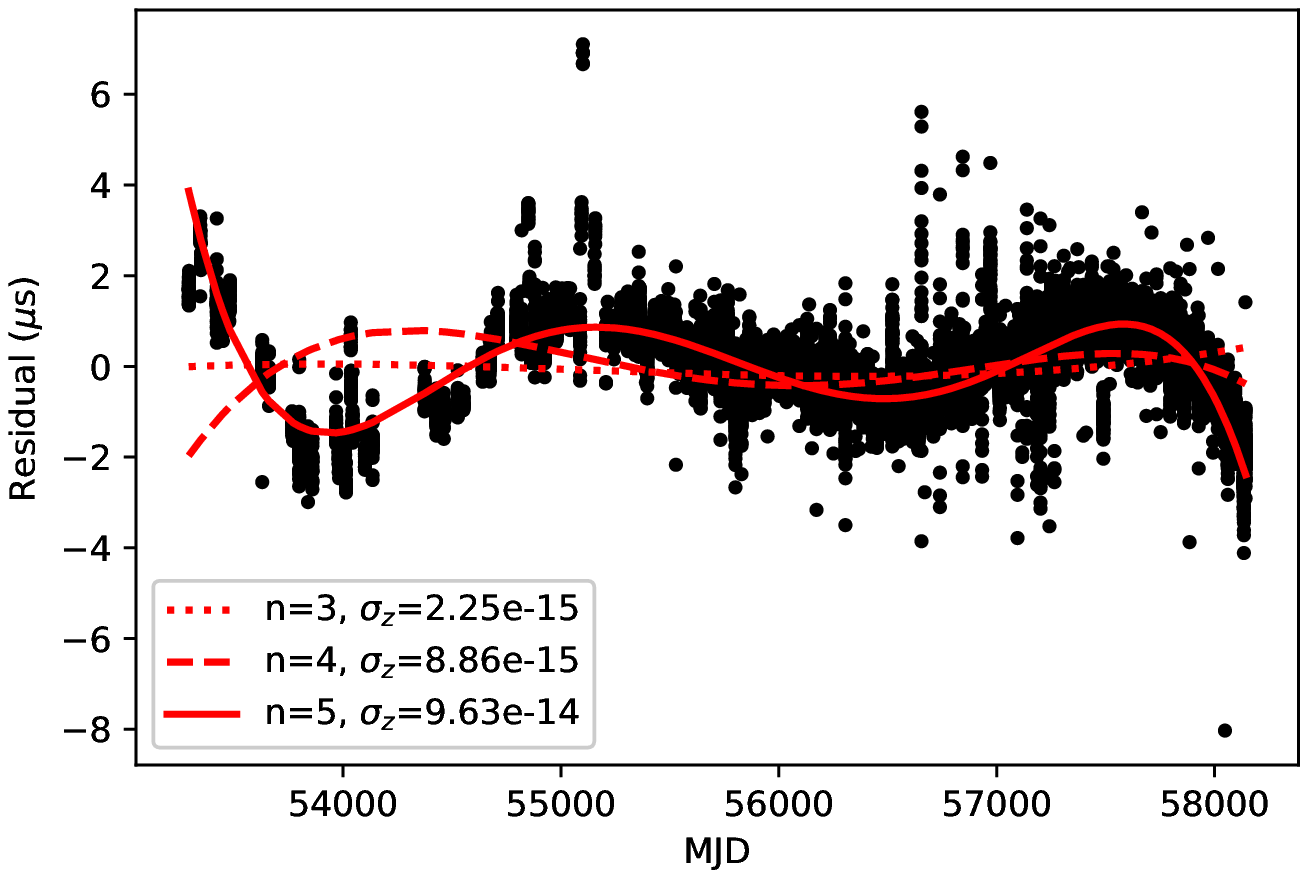}
\caption{The full set of NANOGrav residuals for PSR~B1937+21 exhibits significant red noise variations, and third- or fourth-order polynomials (dotted and dashed lines, respectively) give poor fits; a fifth-order polynomial (solid line) is necessary to obtain a good fit. In each case, the third-order coefficient is used to compute a nominal $\sigma_z$ according to Equation~\ref{eqn-sigmaz}. The result for the fifth-order polynomial yields the open blue square point in Figure~\ref{fig-1937-sigmaz}. \label{fig-nanograv-polytest}}
\end{center}
\end{figure}

The full set of NANOGrav B1937+21 residuals requires special treatment: it is five years longer than the B1937+21 data set used by \cite{Matsakis97}, and a third-order polynomial fails to yield a good fit to the residuals for the largest $\tau$. Figure~\ref{fig-nanograv-polytest} shows third-, fourth-, and fifth-order polynomial fits for this case, and the corresponding $\sigma_z$ calculated from the third-order coefficient of each polynomial fit. The resulting nominal $\sigma_z$ values differ substantially from one another. 

For radio $\sigma_z$, we use the final timing solutions from Section~\ref{sec-residuals}, which include fitted DMX offsets. For X-ray $\sigma_z$, 
we adopt the best-fit astrometric parameters (position, proper motion, parallax) from these final radio solutions---the span of \textit{NICER} TOAs is less than one year, too short to allow robust fitting of these parameters.
ISM propagation parameters (DM, DMXs) are not relevant for X-ray timing. 

\begin{figure}
\begin{center}
\includegraphics[clip,trim={0 0 0 30},width=0.7\textwidth]{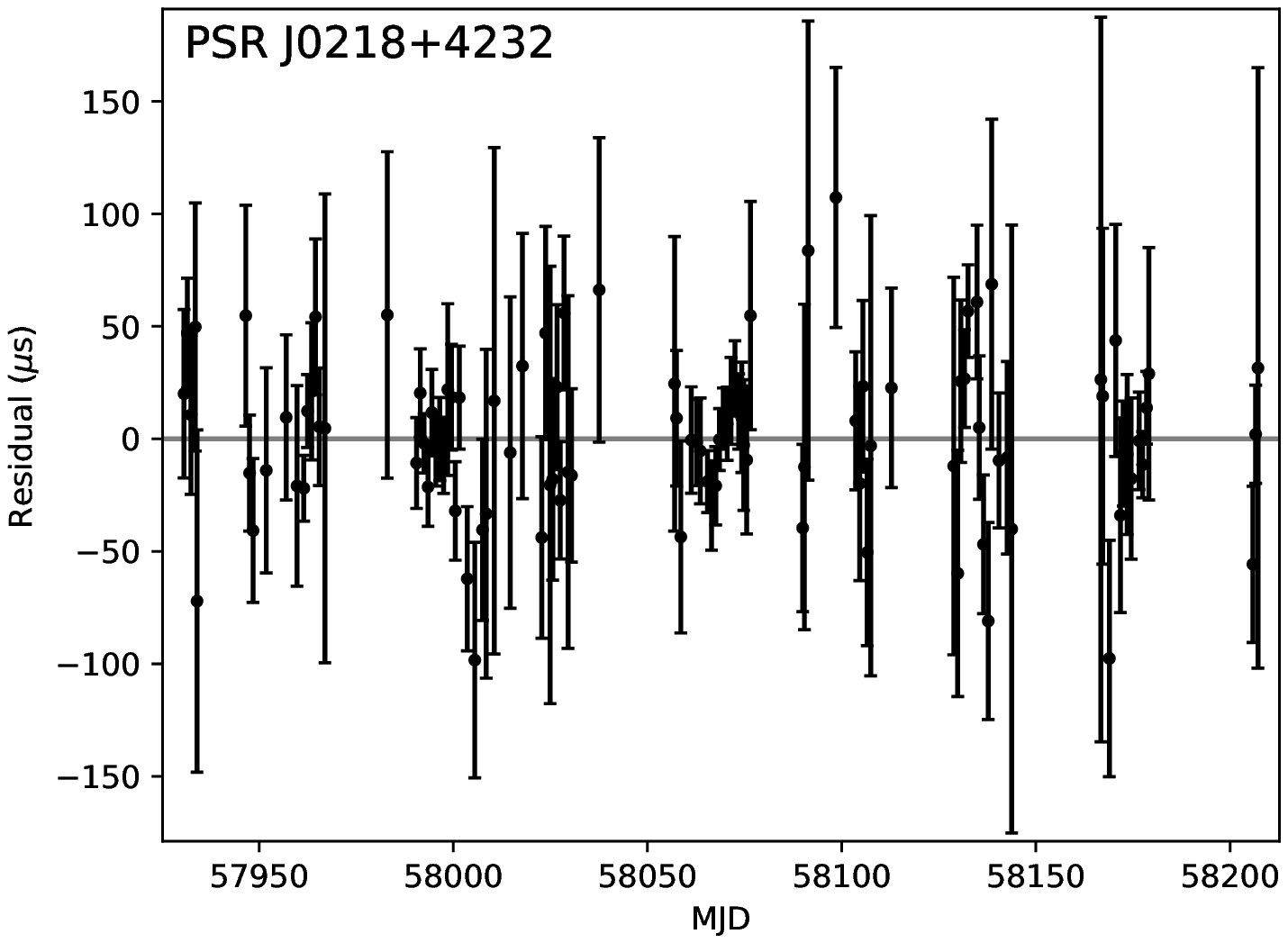} \\
\includegraphics[clip,trim={0 0 0 30},width=0.7\textwidth]{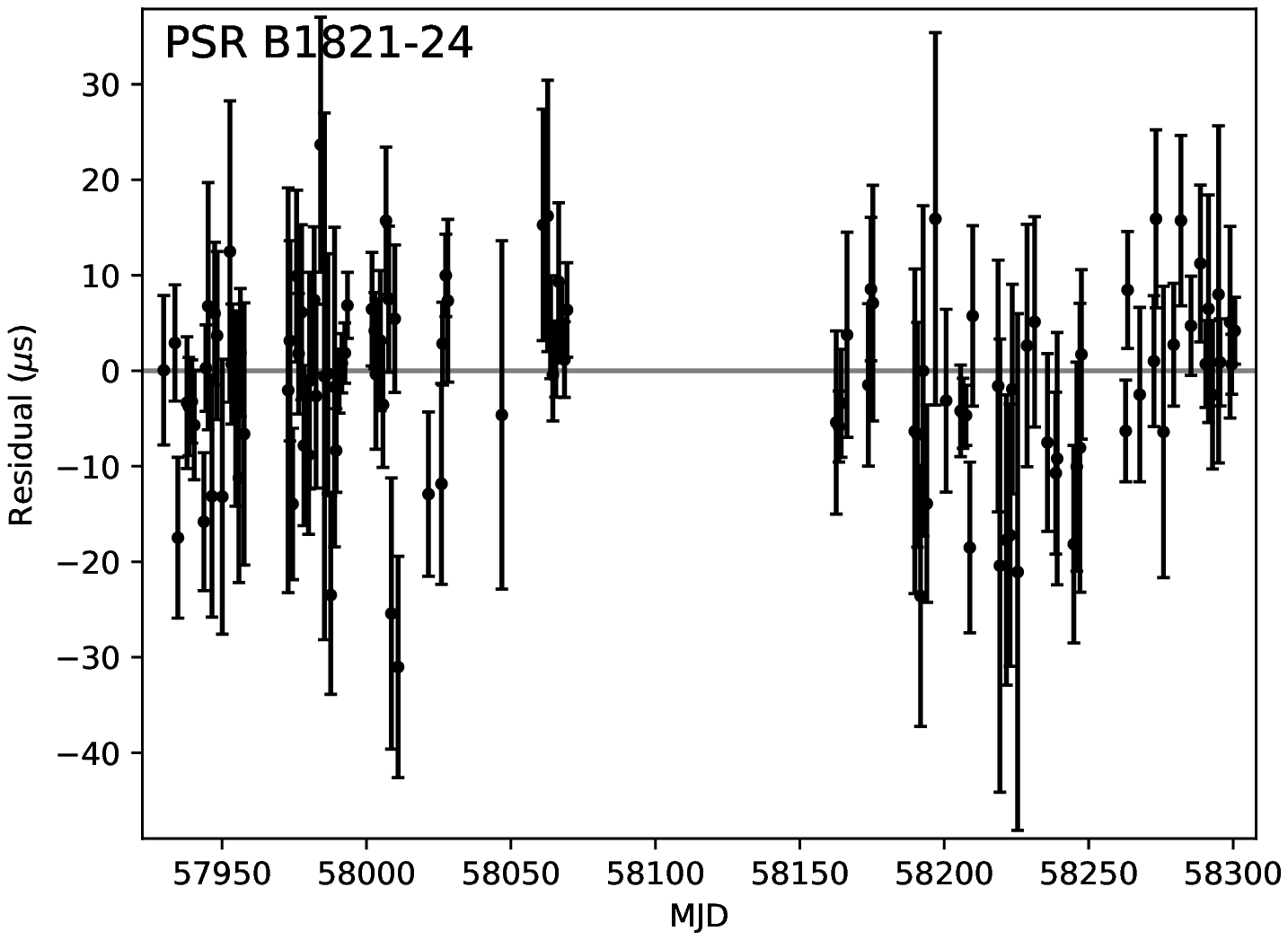} \\
\includegraphics[clip,trim={0 0 0 30},width=0.7\textwidth]{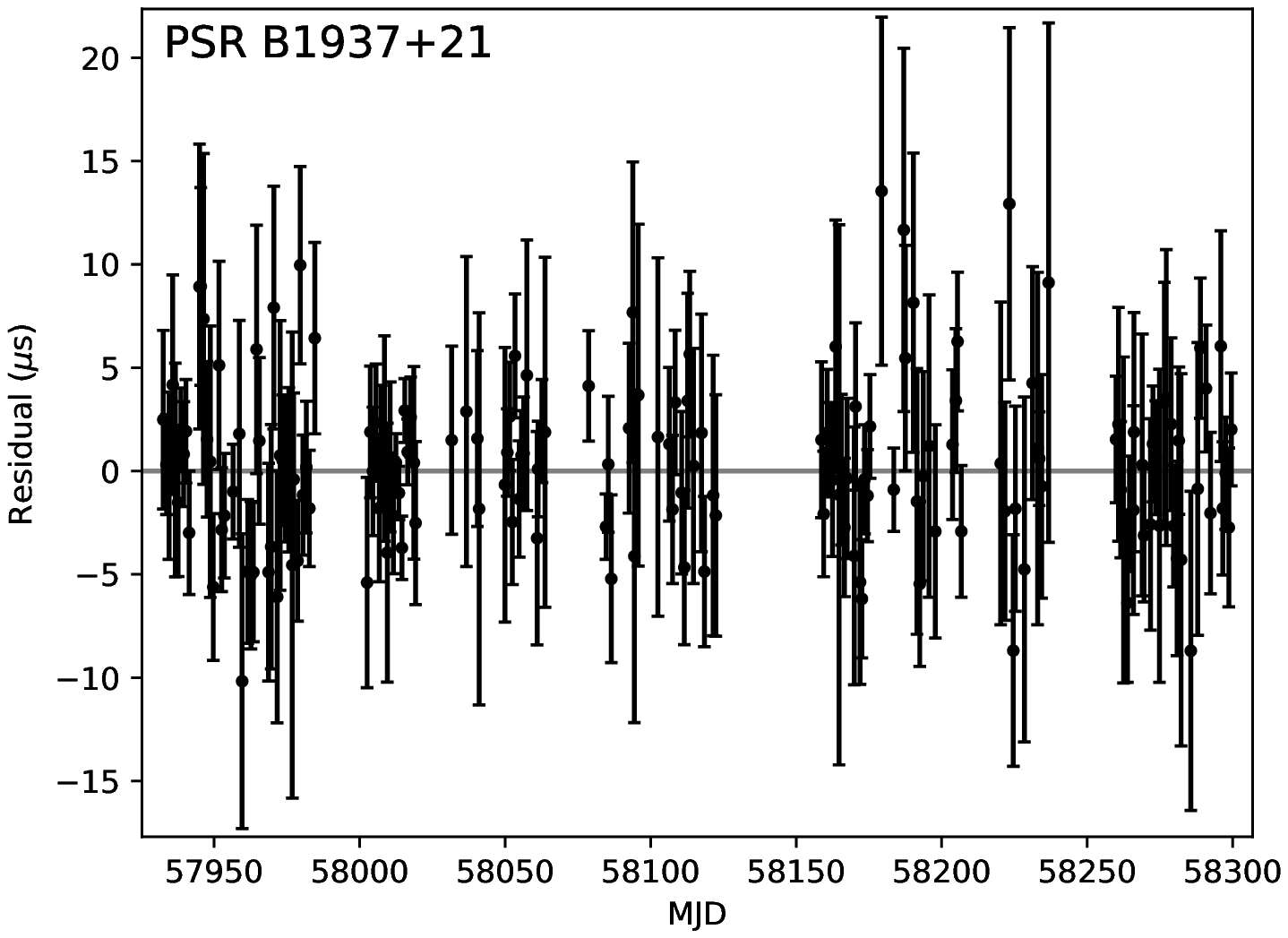}
\caption{\textit{NICER} timing residuals with respect to the final timing solutions after refitting the rotational period, period derivative, and in the case of PSR~J0218+4232, also binary parameters using only \textit{NICER} TOAs, as described in Section~\ref{sec-sigmaz}.\label{fig-nicer-res}}
\end{center}
\end{figure}

We then use {\tt PINT} to refit the rotational periods, period derivatives and, in the case of PSR~J0218+4232, orbital parameters. This yields smaller \textit{NICER} residuals compared to using the radio timing solution without refitting, because by refitting we remove some of the chromatic, imperfectly modeled ISM effects from these parameters that are present in the radio solution and do not apply to \textit{NICER} TOAs. However, the new best-fit rotational period and orbital parameters retain some covariance with the fixed parameters from the initial radio solution, and therefore our updated \textit{NICER} timing solution is not completely free from the influence of propagation effects plaguing the radio data nor from covariances with the astrometric parameters. \textit{NICER} timing residuals with respect to this partially refitted timing solution for each pulsar are plotted in Figure~\ref{fig-nicer-res}. { We include these final X-ray timing solutions for our three pulsars as supplementary electronic materials.}

Figures~\ref{fig-0218-sigmaz}--\ref{fig-1937-sigmaz} show $\sigma_z$ vs.\ the length of non-overlapping intervals $\tau$ over which we fit cubic polynomials to the radio and \textit{NICER} timing residuals of PSR~J0218+4232, PSR~B1821$-$24, and PSR~B1937+21. If residuals are affected only by white noise, $\sigma_z \propto \tau^{-1.5}$ \citep{Matsakis97}; for reference, this case is illustrated by a dashed line in each figure drawn through the leftmost \textit{NICER} point. A solid line shows the weighted fit for \textit{NICER} $\sigma_z$ points, and a grey area shows the fit uncertainty. The redder the noise, the shallower the line slope. 
While the fit to \textit{NICER} $\sigma_z$ points is redder than white noise for all three pulsars, the white noise slope is well within fit uncertainties. The grey region of uncertainty will shrink as we accumulate more \textit{NICER} data. 

For PSR~B1937+21, radio $\sigma_z$ points in Figure~\ref{fig-1937-sigmaz} show a turn-up at $\gtrsim 600$~days in the NRT data set, and at $\gtrsim 1000$~days in the NANOGrav data set. This is roughly consistent with the results of \cite{Matsakis97}, where the turn-up occurs at $\gtrsim 800$~days. At the full length of the NANOGrav B1937+21 data set, $\sim 5000$ days, there is a spurious turn-down if $\sigma_z$ for that time scale is calculated from a third-order fit because a cubic does not describe well the residuals at that time scale (see Figure~\ref{fig-nanograv-polytest}). Therefore, in addition to $\sigma_z$ from the poor cubic fit we show (with an open blue square in Figure~\ref{fig-1937-sigmaz}) $\sigma_z$ from the third-order coefficient of a fifth-order polynomial, which is the best fit to the full set of NANOGrav B1937+21 residuals. 
 
The fits to \textit{NICER} points are also projections of $\sigma_z$ for \textit{NICER} TOAs of the three pulsars by the completion of the first 2~years of the \textit{NICER} mission, denoted by a vertical line in each figure. 
PSR~J0218+4232 has a much wider X-ray pulse profile (Figure~\ref{fig-0218-prof}) than PSR~B1821$-$24 and PSR~B1937+21 (Figures~\ref{fig-1821-prof} and \ref{fig-1937-prof}, respectively), which results in larger TOA uncertainties for the same exposure time per TOA (compare Figure~\ref{fig-0218-crlb} with Figures~\ref{fig-1821-crlb} and \ref{fig-1937-crlb}) and higher $\sigma_z$ for the same timing baseline.

While the effects of white noise on timing precision can be mitigated by increasing the total number of observations as well as the total time span of TOAs, this is not true for red noise. Since both rotational and propagation effects causing red noise are stochastic and slowly-varying, it becomes more prominent on large time scales. In addition, while on short time scales red noise may be subsumed in fitted timing parameters like rotational or DM derivatives, these parameters are stochastic and do not extrapolate well beyond the fitted data set. As a consequence, we expect the slope of $\sigma_z$ to become shallower and eventually level off with increasing $\tau$. Once that limit is reached, accumulating more TOAs no longer results in a lower cumulative RMS timing residual over the entire TOA span. Figure~\ref{fig-1937-sigmaz} shows that this is the case for PSR~B1937+21 in radio for $\tau > 10^3$~days. 

NANOGrav uses Bayesian analysis first to detect whether a pulsar's residuals contain red noise and, if that is the case, to model its amplitude and spectral index (\citealt{9yr}, \citealt{11yr}) along with the values of other timing parameters in the presence of red noise. However, these efforts admit that some of the red noise is still absorbed, in unknown proportions, by the achromatic first period derivative and the chromatic DMX offsets. Since \textit{NICER} pulsar residuals do not contain ISM-dependent chromatic red noise, our analysis can also help separate chromatic, ISM-induced red noise from achromatic red noise due to intrinsic rotational instabilities in existing long-term radio timing data sets. 

Because the \textit{NICER} TOA span is still $\lesssim 1$~year, we do not yet expect to see any leveling off in \textit{NICER} $\sigma_z$ values in Figures~\ref{fig-0218-sigmaz}--\ref{fig-1937-sigmaz}. For PSR~B1821$-$24, the rightmost \textit{NICER} $\sigma_z$ point in Figure~\ref{fig-1821-sigmaz} indicates that, within error bars, there may or may not be a turn-up in \textit{NICER} points for $\tau \gtrsim 300$ days. This point derives from a single $c_3$ value when a cubic polynomial is fitted to the full span of \textit{NICER} residuals. Accumulating more data will allow us to clarify how \textit{NICER} $\sigma_z$ values behave at time scales of hundreds to thousands of days. 


\section{Discussion}
 
One question we want to answer is: how much red noise in MSP timing data is attributable to ISM propagation effects? \textit{NICER} observations do not suffer from the ISM propagation effects and consequent red noise that plagues radio observations. One of the goals of the \textit{NICER} mission is to test whether this leads to $\sigma_z$ leveling off earlier for radio TOAs, or whether intrinsic rotational instabilities dominate, in which case both radio and \textit{NICER} $\sigma_z$ would level off at similar time scales $\tau$. The answer to this question has implications for detecting the stochastic, nanohertz gravitational wave background, a regime accessible to PTAs and complementary to the regime of gravitational wave emission explored by LIGO. Currently, efforts toward detecting nanohertz gravitational waves rely on ground-based radio observations of MSPs and have yet to produce a detection. Obtaining the best possible timing precision for dozens of MSPs is central to these efforts. MSP timing with \textit{NICER} and future X-ray instruments may prove a valuable or perhaps even critical addition to radio PTAs, either by providing a way to better evaluate and mitigate the effects of red noise on radio PTA data sets, or by producing X-ray TOA sets on some PTA MSPs that may be used in conjunction with radio TOAs. 

Radio pulsars are considered good candidates for inclusion in PTAs if TOAs with uncertainty $\lesssim 1$\,$\mu$s can be obtained within integration times of $\lesssim 30$\,minutes. Because X-ray photons are sparse, the \textit{NICER} exposure needed to achieve the same TOA uncertainty is longer. In Figures~\ref{fig-0218-crlb}--\ref{fig-1937-crlb}, the best theoretical match to \textit{NICER} TOA uncertainties vs.\ exposure is the numerical CRLB result based on a smooth two-Gaussian pulse profile, denoted with a solid line. Extrapolating this to a TOA uncertainty of 1\,$\mu$s, we find that the necessary exposure is $\sim$150~ks for PSR~B1821$-$24, $\sim$50~ks for PSR~B1937+21, and on the order of a megasecond for PSR~J0218+4232, due to its wide X-ray pulse profile which makes a precise TOA harder to measure. However, while a radio TOA is produced from a single continuous observation recording radio flux density at a regular sampling time, X-ray TOAs may be computed from data spanning days or even weeks of observations, since the arrival time and therefore the phase of each photon is computed individually. 


\textit{NICER} is a technology pathfinder to future missions such as the proposed \textit{Spectroscopic Time-Resolving Observatory for Broadband Energy X-rays} (\textit{STROBE-X})\footnote{\url{https://gammaray.nsstc.nasa.gov/Strobe-X}} and the \textit{enhanced X-ray Timing and Polarimetry Mission} (\textit{eXTP})\footnote{\url{https://www.isdc.unige.ch/extp}}. We consider the implications of Equation~\ref{eq-simple-gaussian} on future missions. The uncertainty of X-ray TOAs depends on the detector area in addition to integration time and background. For a given source, and for a fixed detector and mirror design, the quantity $\left(\alpha+\beta\right)/\alpha$ in Equation~\ref{eq-simple-gaussian} will be the same no matter how many modules (i.e., mirror and detector) are used to make the measurement, but the rate $\alpha$ in counts/s will scale directly as area, which could be represented as $\alpha = \alpha_0 \left(A / A_0\right)$, where $A_0$ is a reference area (that of \textit{NICER}), $\alpha_0$ is the rate corresponding to that area, and $A$ is the area of a hypothetical array of different size. Thus holding the integration time $T$ constant but increasing the area will result in a more precise measurement, scaling as $1/\sqrt{A}$. However, Equation~\ref{eq-simple-gaussian} could be solved for integration time $T$, replacing $\alpha$ and $\beta$ with rates per unit area, as 
\be
\label{eq-int-time}
T = \left(P w\right)^2 \left(\frac{\alpha_0 + \beta_0}{\alpha_0}\right) \frac{1}{\alpha_0\sigma_T^2}\frac{A_0}{A}.
\ee

Equation~\ref{eq-int-time} shows that the integration time required to achieve precision of $\sigma_T$ scales as $1/A$. To get the integration time for PSR~B1821$-$24 from $\sim$150~ks down to 15~ks would require increasing the array size from that of \textit{NICER} ($\sim$2000~cm$^2$) to $\sim$2~m$^2$, maintaining the \textit{NICER} design. This is less than the 5~m$^2$ area of \textit{STROBE-X}.
The factor by which integration time needs to be reduced depends upon the application. For detection of long-period gravitational waves it is desirable to over-sample the period of the wave, while achieving the needed accuracy, $\sigma_T$, in each measurement.  

It is possible to realize a substantial further benefit in X-ray arrays by increasing the ratio of mirror size to detector size.  The detector radiation background is independent of the mirrors and accounts for 30\% - 50\% of the background in the \textit{NICER} design. (Other backgrounds of interest scale proportionally with mirror size.) \textit{STROBE-X} and \textit{eXTP} would reduce the ratio $(\alpha_0 + \beta_0)/\alpha_0$ by increasing mirror size without increasing detector size or perhaps even decreasing it, while still producing useful TOAs within a reasonable integration time. These points have been illustrated using the special case of Equation~\ref{eq-simple-gaussian} but are valid more generally, as the scalings do not depend on the pulse shape. 

Overall we find that \textit{NICER} is performing as predicted, and we anticipate making more conclusive statements about the comparison between radio and X-ray timing stability in PSR~J0218+4232, PSR~B1821$-$24, and PSR~B1937+21 when we have accumulated an additional year or more of data. Our calculations of $\sigma_z$ have demonstrated a limitation of this method for evaluating rotational stability in the case of very long sets of timing data containing red noise such as the NANOGrav B1937+21 residuals. In our future analyses of rotational stability based on \textit{NICER} and radio timing residuals we anticipate using a maximum-likelihood method for estimating red noise similar to the one adopted by NANOGrav, described in \cite{9yr}. 

\acknowledgments
This work was supported by NASA through the \textit{NICER} mission and the Astrophysics Explorers Program. This research has made use of data and/or software provided by the High Energy Astrophysics Science Archive Research Center (HEASARC), which is a service of the Astrophysics Science Division at NASA/GSFC and the High Energy Astrophysics Division of the Smithsonian Astrophysical Observatory. Portions of this research performed at the Naval Research Laboratory are sponsored by NASA DPR~S-15633-Y. This research has made use of the NASA Astrophysics Data System (ADS) and the arXiv. The NANOGrav Project receives support from NSF Physics Frontiers Center award number 1430284. Pulsar research at UBC is supported by an NSERC Discovery Grant and by the Canadian Institute for Advanced Research. The Arecibo Observatory is operated by the University of Central Florida in alliance with Ana G. M\'{e}ndez-Universidad Metropolitana, and Yang Enterprises. The National Radio Astronomy Observatory and the Green Bank Observatory are facilities of the National Science Foundation operated under cooperative agreement by Associated Universities, Inc. Part of this research was carried out at the Jet Propulsion Laboratory, California Institute of Technology, under a contract with the National Aeronautics and Space Administration. The Nan\c{c}ay Radio Observatory is operated by the Paris Observatory, associated with the French Centre National de la Recherche Scientifique (CNRS). WWZ is supported by the CAS Pioneer Hundred Talents Program and the Strategic Priority Research Program of the Chinese Academy of Sciences, Grant No. XDB23000000.

\facility{\textit{NICER}}
\software{
\texttt{astropy} (\url{https://ascl.net/1304.002}),
\texttt{PINT} (\url{https://github.com/nanograv/pint}),
HEAsoft (\url{https://ascl.net/1408.004}),
\texttt{Tempo} (\url{https://ascl.net/1509.002}),
\texttt{Tempo2} (\url{https://ascl.net/1210.015})

\newpage

\begin{figure}
\begin{center}
\includegraphics[width=0.75\textwidth]{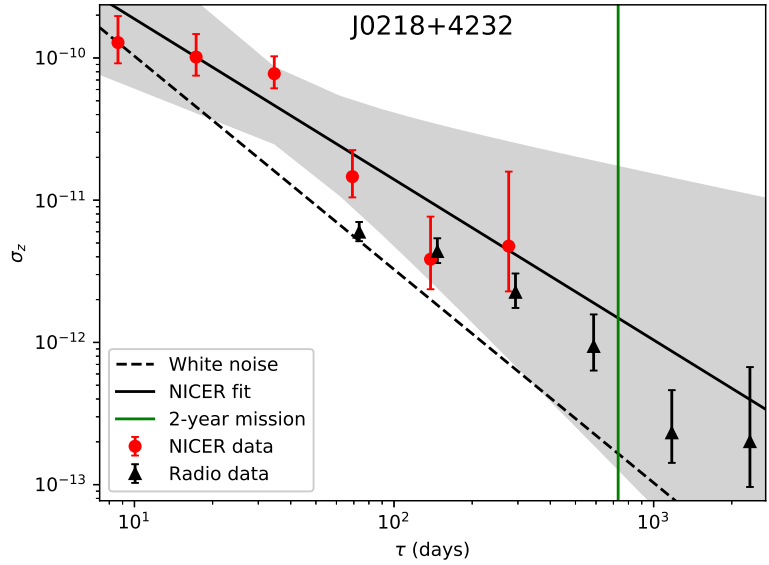}
\caption{A measure of timing stability, $\sigma_z$ (Eqn.~\ref{eqn-sigmaz}), vs.\ the data span over which a third-order polynomial is fitted to the timing residuals of PSR~J0218+4232 according to Section~\ref{sec-sigmaz}. Red and black points show $\sigma_z$ for \textit{NICER} and radio data, respectively. The best fit to the \textit{NICER} points is shown with a solid line and its 1$\sigma$ uncertainty is shaded in grey. A dashed line plotted through the leftmost \textit{NICER} point shows the slope for the case where timing precision is limited by white noise only. A vertical line marks a duration of two years.\label{fig-0218-sigmaz}}
\end{center}
\end{figure}

\begin{figure}
\begin{center}
\includegraphics[width=0.75\textwidth]{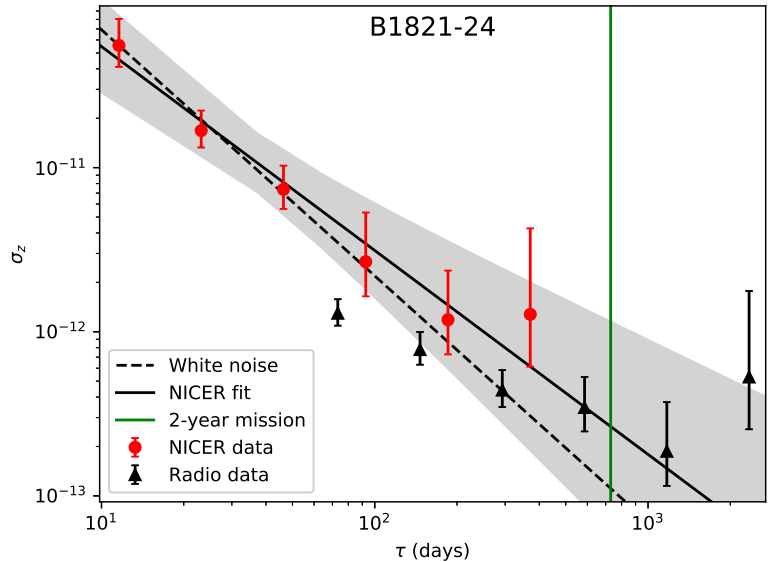}
\caption{Same as Figure~\ref{fig-0218-sigmaz} but for PSR~B1821-24.\label{fig-1821-sigmaz}}
\end{center}
\end{figure}

\begin{figure}
\begin{center}
\includegraphics[width=0.75\textwidth]{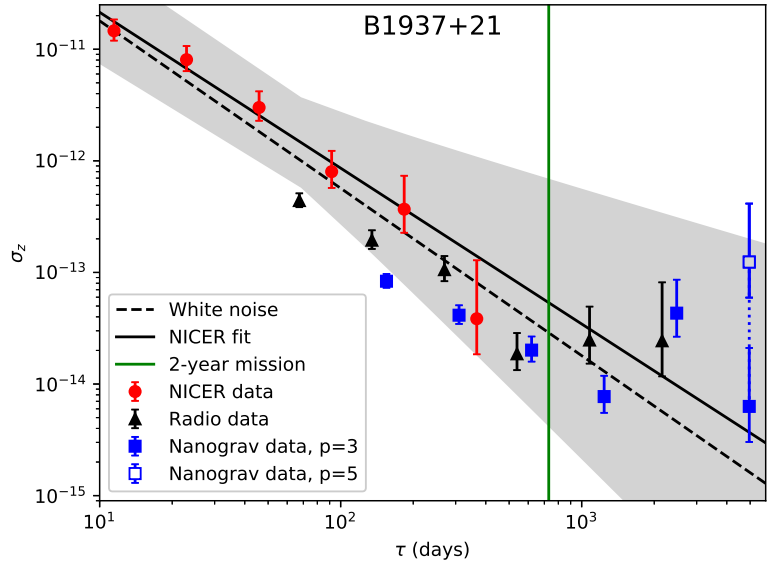}
\caption{Same as Figure~\ref{fig-0218-sigmaz} but for PSR~B1937+21. Red, black, and blue points show $\sigma_z$ for \textit{NICER}, NRT, and NANOGrav data, respectively. Third-order polynomials are used for all data spans $\tau$ and the corresponding $\sigma_z$ values are denoted by filled symbols. A third-order polynomial does not give a good fit to the full span of NANOGrav residuals (Figure~\ref{fig-nanograv-polytest}), and results in a spurious turn-down in the NANOGrav $\sigma_z$ at $\tau \sim 5000$ days. For this case we also show with an open square the nominal $\sigma_z$ value based on the third-order coefficient of a fifth-order polynomial, which is necessary to obtain a good fit to the residuals. 
\label{fig-1937-sigmaz}}
\end{center}
\end{figure}


\clearpage
\begin{thebibliography}

\bibitem[Abbott et al.(2016)]{Abbott16} 
Abbott, B.~P., Abbott, R., Abbott, T.~D., et al.~2016, PhRvL, 116, 061102

\bibitem[Abbott et al.(2017)]{Abbott17} 
Abbott, B.~P., Abbott, R., Abbott, T.~D., et al.~2017, PhRvL, 119, 161101

\bibitem[Arzoumanian et al.(2015)]{9yr}
Arzoumanian, Z. et al. 2015, ApJ, 813, 65

\bibitem[Arzoumanian et al.(2018)]{11yr}
Arzoumanian, Z. et al. 2018, ApJS, 235, 37


\bibitem[Backer et al.(1982)]{Backer82} 
Backer, D.~C., Kulkarni, S.~R., Heiles, C., Davis, M.~M., Goss, W.~M.~1982, Nature, 300, 615

\bibitem[Bogdanov \& Grindlay(2009)]{Bogdanov09} 
Bogdanov, S., \& Grindlay, J.~E.~2009, \apj, 703, 1557

\bibitem[Cordes(2013)]{Cordes13}
Cordes, J. M. 2013, 2013, CQGra, 30, 4002

\bibitem[Cusumano et al.(2004)]{Cusumano04}
Cusumano, G. et al. 2004, NucPhysBProcSupp, 132, 596

\bibitem[de Jager et al.(1989)]{deJager89}
de Jager, O. C., Raubenheimer, B. C. \& Swanepoel, J. W. H. 1989, A\&A, 221, 180

\bibitem[de Jager \& B\"{u}sching(2010)]{deJager10}
de Jager, O. C. \& B\"{u}sching, I. 2010, A\&A, 517, 9

\bibitem[Demorest(2007)]{Demorest07}
Demorest, P. B. 2007, PhD thesis, Univ. California

\bibitem[Devroye(1986)]{Devroye86}
Devroye, L. 1986, Non-Uniform Random Variate Generation, New York: Springer-Verlag

\bibitem[DuPlain et al.(2008)]{DuPlain08}
DuPlain, R., Ransom, S., Demorest, P., et al. 2008, Proc. SPIE, 7019, 70191D

\bibitem[Ferris \& Saunders(2004)]{Ferris04}
Ferris, R. H. \& Saunders, S. J. 2004, Experimental
Astron., 17, 269

\bibitem[Ford et al.(2010)]{Ford10}
Ford, J. M., Demorest, P., \& Ransom, S. 2010, Proc. SPIE, 7740, 0

\bibitem[Gendreau \& Arzoumanian(2017)]{Gendreau17}
Gendreau, K. \& Arzoumanian, Z. 2017, Nature Astronomy, 1, 895
  
\bibitem[Gendreau et al.(2016)]{Gendreau16}
Gendreau, K., Arzoumanian, Z., Adkins, P. W. et al. 2016, SPIE Proc. Vol. 9905, Space Telescopes and Instrumentation 2016: Ultraviolet to Gamma Ray, 99051H

\bibitem[Golshan \& Sheikh(2007)]{Golshan07}
Golshan, A. R. \& Sheikh, S. I. 2007, Proc. of the 63rd Annual Meeting of The Institute of Navigation, p. 413–422

\bibitem[Gotthelf \& Bogdanov(2017)]{Gotthelf17} 
Gotthelf, E.~V., \& Bogdanov, S.~2017, \apj, 845, 159

\bibitem[Guillemot et al.(2012)]{Guillemot12} 
Guillemot, L., Johnson, T.~J., Venter, C., et al.~2012, \apj, 744, 33

\bibitem[Guillemot et al.(2016)]{Guillemot16}
Guillemot, L., Smith, D. A., Laffon, H. et al. 2016, A\&A, 587, 109


\bibitem[Johnston et al.(2013)]{Johnston13}
Johnston, T. J. et al. 2013, ApJ, 778, 2

\bibitem[Kaspi et al.(1994)]{Kaspi94}
Kaspi, V., Taylor, J. H. \& Ryba, M. F. \apj, 428, 713

\bibitem[Knight et al.(2006)]{Knight06}
Knight, H. S., Bailes, M., Manchester, R. N., Ord, S. M. \& Jacoby, B. A. 2006, ApJ, 640, 941

\bibitem[Kuiper et al.(2002)]{Kuiper02} 
Kuiper, L., Hermsen, W., Verbunt, F., Ord, S., Stairs, I., Lyne, A.~2002, \apj, 577, 917


\bibitem[Lorimer \& Kramer(2012)]{Handbook}
Lorimer, D. R. \& Kramer, M. 2012, Handbook of Pulsar Astronomy, Cambridge University Press

\bibitem[Lyne et al.(1987)]{Lyne87} 
Lyne, A.~G., Brinklow, A., Middleditch, J., Kulkarni, S.~R., Backer, D.~C. 1987, Nature, 328, 399

\bibitem[Lyne et al.(2010)]{Lyne10}
Lyne, A. G. et al. 2010, Science, 329, 408

\bibitem[Manchester et al.(2013)]{Manchester13}
Manchester, R. N., Hobbs, G., Bailes, M. et al. 2013, PASA, 30, 17

\bibitem[Matsakis et al.(1997)]{Matsakis97}
Matsakis, D. N., Taylor, J. H. \& Eubanks, T. M. 1997, A\&A, 326, 924

\bibitem[Mineo et al.(2000)]{Mineo00} 
Mineo, T., Cusumano, G., Kuiper, L., Hermsen, W., Massaro, E., Becker, W., Nicastro, L., Sacco, B., Verbunt, F., Lyne, A.~G., Stairs, I.~H., Shibata, S.~2000, A\&A, 355, 1053

\bibitem[Mitchell et al.(2018)]{Mitchell18}
Mitchell, J. W. et al. 2018, Proc. of the 15th International Conference on Space Operations, American Astronautical Society

\bibitem[Navarro et al.(1995)]{Navarro95} 
Navarro, J., de Bruyn, A. G., Frail, D. A., Kulkarni, S.~R., Lyne, A.~G.~1995, \apj, 455, L55

\bibitem[Ng et al.(2014)]{Ng14} 
Ng, C.-Y., Takata, J., Leung, G.~C.~K., Cheng, K.~S., Philippopoulos, P.~2014, \apj, 787, 167

\bibitem[Nicastro et al.(2004)]{Nicastro04} 
Nicastro, L., Cusumano, G., L\"ohmer, O., Kramer, M., Kuiper, L., Hermsen, W., Mineo, T., Becker, W.~2004, A\&A, 413, 1065

\bibitem[Prigozhin et al.(2016)]{Prigozhin16}
Prigozhin, G., Gendreau, K., Doty, J. P. et al. 2016, SPIE, 9905, 11

\bibitem[Ray et al.(2008)]{Ray08}
Ray, P. S., Wolff, M. T., Demorest, P. et al. 2008, 40 YEARS OF PULSARS: Millisecond Pulsars, Magnetars and More, AIP Conference Proceedings, 983, 157

\bibitem[Ray et al.(2011)]{Ray11}
Ray, P. S., Kerr, M., Parent, D. et al. 2011, ApJS, 194, 17


\bibitem[Rots et al.(1998)]{Rots98} 
Rots, A.~H., Jahoda, L., Macomb, D.~J., et al.~1998, \apj, 501, 749

\bibitem[Rutledge et al.(2004)]{Rutledge04} 
Rutledge, R. E., Fox, D. W., Kulkarni, S. R., Jacoby, B. A., Cognard, I., Backer, D. C., \& Murray, S. S. 2004, \apj, 613, 522

\bibitem[Saito et al.(1997)]{Saito97} 
Saito, Y., Kawai, N., Kamae, T., Shibata, S., Dotani, T., Kulkarni, S.~R.~1997, \apj, 477, L37

\bibitem[Shannon \& Cordes(2017)]{Shannon17}
Shannon, R. M. \& Cordes, J. M. 2017, MNRAS, 464, 2075

\bibitem[Takahashi et al.(2001)]{Takahashi01} 
Takahashi, M., Shibata, S., Torii, K., Saito, Y., Kawai, N., Hirayama, M., Dotani, T., Gunji, S., Sakurai, H., Stairs, I.~H., Manchester, R.~N.~2001, \apj, 554, 316

\bibitem[Taylor(1991)]{Taylor91}
Taylor, J. H. 1991, IEEE, 79, 1054

\bibitem[Taylor(1992)]{Taylor92}
Taylor, J. H. 1992, RSPTA, 341, 117

\bibitem[Webb et al.(2004a)]{Webb04a} 
Webb, N.~A., Olive, J. -F., \& Barret, D. 2004, A\&A, 417, 181

\bibitem[Webb et al.(2004b)]{Webb04b} 
Webb, N.~A., Olive, J.-F., Barret, D., Kramer, M., Cognard, I., L\"ohmer, O.~2004, A\&A, 419, 269

\bibitem[Winternitz et al.(2016)]{Winternitz16}
Winternitz, L.~M.~B., Mitchell, J.~W., Hassouneh, M.~A., et al. 2016, in 2016 IEEE Aerospace Conference, 1-11

\bibitem[Winternitz et al.(2018)]{Winternitz18}
Winternitz, L.~M.~B. et al. 2018, Proc. of the 15th International Conference on Space Operations, American Astronautical Society

\bibitem[Zavlin(2006)]{Zavlin06} Zavlin, V.~E. 2006, \apj, 638, 951

\end{thebibliography}
\end{document}